\documentclass[french,english,,nocjs]{cjs-rcs-article}
\usepackage[utf8]{inputenc}
\usepackage[T1]{fontenc}

\usepackage{graphicx} % Required for inserting images
\usepackage{booktabs}
\usepackage[utf8]{inputenc}
\usepackage[T1]{fontenc}
\usepackage{helvet}
\usepackage{makecell}
\usepackage{amsmath}
\usepackage{multirow}
\usepackage{overpic} % For the overpic environment
\usepackage{amsfonts} 
\usepackage{enumitem}
\usepackage{dsfont}

\usepackage{float}
\usepackage{mathrsfs} % For calligraphic math fonts with bold support
\usepackage{caption}
\usepackage{lipsum}
\usepackage{natbib}
\usepackage{amsthm}
\usepackage{diagbox}
\usepackage{algorithm}
\usepackage{algorithmic}
\usepackage{empheq}
\usepackage{subcaption}
\usepackage{newunicodechar}
\usepackage[normalem]{ulem}
\usepackage{xcolor}
\newunicodechar{́}{\'}
\usepackage[autostyle=true]{csquotes}

 \title{Detection of Structural Distortions in Functional Time Series}
 \author[email=dbnjanad@gmail.com]
{Debanjana \surname{Datta}}
 \affil{Applied Statistics Unit, Indian Statistical Institute, Bangalore, India}
 \author{Rituparna \surname{Sen}}
 \affil{Applied Statistics Unit, Indian Statistical Institute, Bangalore, India}
 \author{Nalini \surname{Ravishanker}} 
 \affil{Department of Statistics, University of Connecticut, Storrs, USA}
 
 \begin{englishabstract}

In the era of modern data science, the rapid proliferation of high-dimensional and functional datasets has fostered increasing interest in the investigation of paradigm shifts and structural breaks.
 Unlike classical univariate time series, structural changes in functional data need not occur simultaneously across the entire domain; instead, they may emerge locally, producing heterogeneous distortions across the underlying functional structure.
The patterns of instability often exhibit sparsity, where it is not known \textit{a priori} which specific parameters are undergoing a transition. However, in functional contexts, these shifts are often "localised". The difficulty lies in the high dimensionality of the parameter space, where the signal-to-noise ratio may be low for individual components, necessitating the aggregation of information across dimensions to detect a global change.
This paper addresses the problem of detecting structural shifts in a functional time series from a Bayesian perspective. 
We have developed various novel methodologies that capture the inherent structural distortion in a sequence of random functions, both individually and simultaneously. The formulation of the problem is based on the state-space representation of a functional time series. Efficient Blocked Gibbs Sampling algorithms have been proposed to identify these locations accurately. Further, we demonstrate the effectiveness of our methods on several financial and temperature datasets.

 \end{englishabstract}

 \begin{frenchabstract}
À l'ère de la science des données moderne, la prolifération de jeux de données fonctionnels et de grande dimension a suscité un intérêt pour l'étude des changements de régime ou des ruptures structurelles. Contrairement aux séries temporelles univariées classiques, dans le cas des données fonctionnelles — où les observations peuvent être considérées comme des courbes ou des surfaces continues — une altération structurelle ne se manifeste pas nécessairement simultanément sur l'ensemble de l'opérateur fonctionnel.
Au contraire, les phénomènes d'instabilité présentent souvent un caractère de parcimonie, sans que l'on sache *a priori* quels paramètres spécifiques subissent une transition. Toutefois, dans un contexte fonctionnel, ces changements sont souvent « localisés » au sein de l'espace de Hilbert des fonctions. La difficulté réside dans la grande dimensionnalité de l'espace des paramètres, où le rapport signal sur bruit peut être faible pour les composantes individuelles, nécessitant ainsi une agrégation de l'information à travers les dimensions pour détecter un changement global.
Cet article aborde le problème de la détection de points de rupture dans des séries temporelles fonctionnelles selon une approche bayésienne.
Nous avons développé plusieurs méthodologies inédites capables de saisir l'altération structurelle inhérente à une séquence de fonctions aléatoires, tant individuellement que simultanément. La formulation du problème repose sur une représentation en espace d'état de la série temporelle fonctionnelle. Des algorithmes efficaces d'échantillonnage de Gibbs par blocs ont été proposés pour identifier correctement ces points de rupture. Enfin, nous illustrons l'application de ces méthodes à l'aide de jeux de données financiers.
 
 \end{frenchabstract}


\begin{document}

%% Title page (MANDATORY - do not delete the following line!)
\maketitle % typeset the title page
%% Numbered introduction (MANDATORY)

\section{INTRODUCTION}
\label{sec:introduction} % example of a section label

The growing utility of Functional Data Analysis (FDA) stems from its ability to process and characterize complex, continuous functional phenomena in diverse domains such as biomedicine, finance, and atmospheric sciences. A significant methodological limitation is the frequent and often unrealistic assumption of independent and identically distributed observations. For observational data, this assumption is regularly violated; specifically, the time-evolving nature of the underlying stochastic process can induce structural shifts within the functional data. Consequently, developing methods to model these temporal dependencies is an area of paramount importance. Hence, covariance estimation, along with mean estimation, is a fundamental problem in the domain of functional data and has multifaceted applications, for example, in regression, prediction, classification, etc. Nonparametric estimation of covariance kernels is still an active field of research in functional data analysis (see (\cite{wang2016functional}) for a complete overview). In functional data, our observations are random continuous trajectories generated by a stochastic process; hence, they are ultra-high-dimensional objects. Generally, the observations are allowed to vary on both sparse and dense support. In practice, these functions are observed on a finite number of grid points in a compact domain. Examples of such functions include ECG data, concentration of certain proteins in blood(\cite{kokoszka2017introduction}), and others. The most common approach used to model functional observations is to project them onto some finite-dimensional space using some dimension reduction technique, namely Functional Principal Component Analysis (FPCA). \cite{ramsay2005functional} has enhanced the FDA literature with detailed discussions on several techniques and the usefulness of FPCA. To avoid potential loss of information inherent in projecting functional data onto a finite set of principal component directions, we bypass dimension reduction and instead propose a methodology that operates directly on the entire function space. Direct analysis preserves the full variation space and avoids tuning parameter dependencies (e.g., basis rank $p$, smoothing parameter $\lambda$). However, operating without pre-smoothing increases sensitivity to discrete measurement noise and limits standard quadrature when data are sparsely observed. The state-space formulation naturally allows for sparse observations with noise. In functional spectral analysis, pre-smoothing can similarly be bypassed by working directly with the spectral density operator \cite{bagchi2018simple}. There is an abundance of literature that models functional data using a regression approach (see \cite{faraway1997regression}, \cite{ramsay2005functional}, etc.).

A functional time series (FTS) arises when we observe a random function at each time point, generally on a dense support. Examples of such functions include daily stock price curves, yield curves(\cite{sen2019time}), and others. In a multivariate setting, along with serial correlation, one needs to study the correlation among multiple functions. In this context, \cite{gabrys2007portmanteau} devised a test to verify the independence and identical distribution assumption of functional observations. The functional autoregressive (FAR) model, developed by \cite{bosq2000linear}, plays a central role in modeling and predicting functional time series. \cite{aue2015prediction} have further enriched the domain of prediction methodologies for FTS.

Dependent functional time series can be modeled in several ways depending on the nature of the temporal dependence, the complexity of the dynamics, and the underlying application. Some of the most widely used models are Functional factor models and functional generalised autoregressive conditional heteroscedasticity (fGARCH) models. Functional factor models (see \cite{hallin2023factor}, \cite{kowal2023semiparametric}) provide an efficient representation of dependent functional time series by expressing each functional observation as a linear combination of a small number of latent factors. Functional factor models are particularly useful for high-dimensional functional datasets because they reduce the dimensionality of the observations while preserving the underlying temporal dependence. They have found wide applications in economics, finance, environmental sciences, and forecasting of functional time series. While functional factor models primarily focus on modelling the conditional mean structure of a functional time series, they are often inadequate for capturing time-varying volatility. To address this limitation, functional generalised autoregressive conditional heteroscedasticity (fGARCH) models have been developed as functional extensions of the classical GARCH framework. The fGARCH model (see \cite{aue2017functional}, \cite{li2025functional}) allows the conditional variance structure to evolve over time and captures volatility clustering, a phenomenon commonly observed in financial and economic data. Consequently, functional GARCH models are particularly useful for analyzing high-frequency financial curves, intraday return trajectories, yield curves, and other functional datasets exhibiting heteroscedastic behavior.

In practical cases, it is often important to verify whether a single model works well for the entire data set. Since the underlying stochastic nature of a process evolves over time, accounting for these structural transitions is imperative to safeguard against the risk of erroneous inference. Change point analysis is the systematic process of detecting shifts with respect to any characteristic of the data or even in the data-generating mechanism. Notable works on a single change point problem include \cite{hinkley1971inference}, \cite{pettitt1979non} and recently by \cite{jewell2022testing}, \cite{dalla2026testing} and others. The corresponding multiple change point framework has been thoroughly studied in \cite{hawkins2001fitting}, \cite{lavielle2006detection}, \cite{matteson2014nonparametric}, and others. 

The study of change point in functional data analysis include the test for the equality of functional autoregressive operators (FAR) proposed by \cite{horvath2010testing} and the test for the stability of the volatility and cross-volatility structure of a functional process developed by \cite{aue2009break}.We shall deal with the estimation of the operator, whereas \citet{horvath2010testing} dealt with the testing problem—typically quantifying the action of the operator on the space spanned by the $p$ most important functional principal component (FPC) directions of the entire dataset. Additionally, the constant mean assumption in a sequence of functional observations was tested by \cite{berkes2009detecting}. At the same time, \cite{aue2009estimation} also contributed to the work dealing with the precise position of such a change. In contrast, \cite{banerjee2018more} provided a different test for detecting the change point in the mean in a sequence of independent and weakly dependent functional data.

The change-point problem is well studied in the Bayesian context as well. Pioneering works have been done by \cite{chernoff1964estimating}, \cite{smith1975bayesian}, \cite{carlin1992hierarchical}, \cite{barry1993bayesian} etc. The Bayesian treatment of change-point problems can be viewed from a different perspective, as in \cite{cappello2025bayesian}, where they use a variational approximation to the model posterior distributions. In functional data, recent work has been done by \cite{li2021bayesian}, where the setup is independent, and a change point is detected where the ratio of mean similarity between groups and mean similarity within groups is minimum. Further, a criterion has been provided for the detection of multiple change points, which is thereby compared with the E-Divisive procedure of \cite{matteson2014nonparametric}.

Apart from this, another broader classification of change point problems lies in its detection methodology, namely the offline and the online schemes. All the works cited earlier are on offline mode, a retrospective study. The online detection methodology, i.e., the prospective study, has been introduced by \cite{page1954continuous} and \cite{wald1992sequential}. Bayesian methodology for online detection is currently a budding area of research, while its inception goes back to the pioneering work of  \cite{fearnhead2007line}. Substantial development in this area has been made by \cite{Yiğiter03072015}, \cite{tsaknaki2025bayesian}, and others.

This paper proposes a two-level Gaussian hierarchical model for detecting structural shifts in a functional time
series. The very formulation of the model will overcome the issue of inconsistent solutions arising in the FPCA approach and produce efficient results. We represent the hierarchical model as a dynamic linear model (DLM) for inferential purpose. Further, observational equations handle discretised functional data with potential measurement error. The evolution equations take into account the correlation among the functional time series and define the process model. A dynamic functional factor model further specifies the dynamic innovation process. We thereby employ an efficient Blocked Gibbs sampling algorithm for estimation purposes. Our proposed methodology is based on a parametric offline scheme. The first major contribution of this article lies in the detection of structural shifts in the underlying latent variable process. Identifying such changes presents a fundamentally challenging problem, as the latent process is not directly observable and its structural dynamics must therefore be inferred from the observed data. To the best of our knowledge, the detection of structural shifts specifically within a latent variable process has not been systematically addressed in the existing literature, primarily due to the inherent complexity associated with latent structures, their indirect observability, and the propagation of structural changes through the observed process. The proposed framework provides a novel perspective for identifying such hidden structural alterations and thereby extends the scope of structural break analysis beyond directly observable processes. The second major contribution concerns the development of a multi-faceted framework for detecting structural changes across multiple parameters simultaneously. In complex data-generating mechanisms, a structural shift may not be confined to a single parameter; instead, it may manifest through changes in several components of the underlying model, potentially occurring at different magnitudes and locations. Consequently, jointly identifying and characterizing such parameter-specific changes constitutes a substantially more challenging problem than conventional single-parameter change-point detection. The proposed approach addresses this complexity by enabling the detection of structural changes across multiple dimensions of the parameter space, thereby providing a more comprehensive characterization of the underlying regime shifts. This multi-parameter perspective represents a novel and important contribution to the literature on structural change detection.

The rest of the paper is organised as follows. In Section \ref{Model}, we shall illustrate the FAR model and its representation in a dynamic linear model (DLM) form, which is further used for inferential purposes. The various notions of change point models have been described in this section using concrete, relevant examples. Section \ref{Prior} provides insights into the prior assumptions of the parameters. We have discussed the initialization procedure of the MCMC as well as various full conditional distribution of the parameters have been derived here. Detection Algorithms as well as the simulation results are provided in Section \ref{algo} and \ref{sim}, respectively. Empirical analysis have been carried out in Section \ref{dataanalysis}. We end with a discussion in Section \ref{conc}.

\section{DYNAMIC LINEAR MODEL FOR FUNCTIONAL TIME SERIES WITH STRUCTURAL BREAKS} \label{Model}

In Section 2.1, we motivate and review methods for modelling a FTS. 

In statistical modelling and time series analysis, assuming that a single process governs an entire observed record over a long horizon is unrealistic.
The September 11, 2001 terrorist attacks (9/11) had a profound impact on global financial markets. \cite{lavielle2006detection} identified a significant regime shift in the daily log returns of Citibank (City), Bank of America (BoA) and Banque Nationale de Paris (BNP) over the period from April 8, 1999, to April 2002. So,
 when the underlying data-generating mechanism undergoes structural modifications due to external or intrinsic shifts, the stochastic properties of the system change. Failure to model these transitions explicitly leads to severe model mis-specification and invalid statistical inference. 

To address this issue, it is necessary to allow for the possibility that the probabilistic structure of the observed series changes at one or more unknown time points. Such changes, commonly referred to as \emph{change points} or \emph{structural breaks}, partition the observation horizon into distinct regimes within which the process remains approximately stationary. By explicitly incorporating these regime shifts into the model specification, it is possible to capture abrupt alterations in the underlying dynamics and obtain more reliable estimation and forecasting. Depending on the aspect of the process that is affected, the distortion may occur in the mean level, the error variance, the temporal dependence mechanism, or a combination thereof. Therefore, it is of utmost importance to propose a model that explicitly accounts for the relevant changes.

A detailed description of our proposed models with structural breaks is provided in Section 2.2.

\subsection{\MakeUppercase{Hierarchical Modelling}} \label{hmodel}

%\textcolor{red}{Since we are not using factor models or fGARCH models, maybe the next 3 paras can also move into the Intro? I have commented out here. Decide and maybe move? }\textcolor{blue}{resolved}

We consider the functional autoregression (FAR) framework for modeling functional time series. Let $Y_{1}$,\ldots,$Y_{T}$ be a time-ordered stationary sequence of random functions in \ $L^{2}(\mathcal{T})$, where $\mathcal{T}$ is a compact indexing set. For simplicity, we consider $\mathcal{T}$=[0,1].

%\textcolor{red}{Below, do you mean the mean of the SP $Y_t$?}\textcolor{blue}{(resolved) } 

%\textcolor{red}{Is there a better notation for $\Psi(\alpha)(k)$? Two brackets?} \textcolor{blue}{(Standard Notation) } 
%

%\textcolor{red}{Shd u clarify the relation between $Y_1,\ldots, T_T$ and $Y_1(u),\ldots, Y_T(u)$?} \textcolor{blue}{resolved}

%\textcolor{red}{Also, I dont see you using $Y_t(u)$ or $]\alpha_t(u)$ anywhere below in this section?}\textcolor{blue}{See equations (1) and (3)}

For any $u$ $\in$ $\mathcal{T}$, let $Y_t(u)$ denote the value of the function $Y_t$ at $u$. Define
\begin{math}
 \mu(u)=E(Y_1(u))
\end{math} and
\begin{align}
 \alpha_t(u) = Y_t(u)-\mu(u)
\end{align}
be the mean centered functional process.

%\textcolor{red}{FAR models have been defined before 2017, as u show in the intro. Do u mean u are using the 2017 notations? }\textcolor{blue}{Yes}

Following the notation in \cite{kokoszka2017introduction}, we write a Functional Autoregressive Process of order 1 (FAR(1)) as 
\begin{align*}
\alpha_{t}=\Psi (\alpha_{t-1})+\epsilon_{t},~t = 1,\ldots, T 
\end{align*}
%\begin{math}
 % \alpha_{t}=\Psi (\alpha_{t-1})+\delta_{t}\
%\end{math};\ t = 1,\ldots, T
where
\begin{math}
\Psi : L^{2} \rightarrow L^{2}
\end{math} is an autoregressive operator and 
\begin{math}
\{\epsilon_{t}\}_{t\leq T}
\end{math}
is a sequence of independently and identically distributed (i.i.d.) random errors in $L^{2}$ . For interpretability, we restrict our attention only to integral operators defined by 
\begin{math}
 \Psi(\alpha)(k)= \int_{0}^{1} \psi (k,s) \alpha(s)\,ds 
\end{math}, 
where the bivariate functions $\psi(k, s)$ are called the kernels of the Hilbert--Schmidt operator $\Psi$. The assumption that $\int_{0}^{1} \int_{0}^{1} \psi^2(k, s) \,dk \,ds < 1$ guarantees stationarity of the $\{\alpha_t\}_{t=1}^{T}$ process.

%\textcolor{red}{I am editing the writing below: are $v_t$ and $\nu_t$ the same or diff? It is unclear in this writing. Also, v and nu look so alike, people may have the same doubt I had, that maybe it is a typing mistake. will a diff symbol be helpful?}\textcolor{blue}{resolved}

The functions $\{Y_{t}\}_{t=1}^{T}$ are not observed directly, but %rather with potential 
include a measurement error, i.e.,
%Suppose we observe 
\begin{align}
 y_{t}=Y_{t}+ \nu_{t},~ t = 1,\ldots,T
 \end{align}
where 
\begin{math}
y_{t} \in L^{2}
\end{math} is sampled 
from 
$Y_{t}$ with noise $\nu_{t}$, and 
$\nu_{t}$ is a white noise process independent of $\epsilon_t$.
Although the underlying process may follow a specific model, the corresponding observation process need not adhere to the same modelling framework. %\\
Proposition 1 of \cite{kowal2019functional} illustrates that $y_{t}$ follows FARMA(1,1) even when $\alpha_t$ follows FAR(1). Thus, in situations when observations come with noise, considering a FAR structure for them leads to a model mis-specification. 

To mitigate the effects of model mis-specification and the consequent risk of erroneous statistical inference, we propose a two-level hierarchical modelling framework. The first level, or the evolution level, is designed to capture the latent dependence structure underlying the data-generating mechanism, thereby accounting for the intrinsic relationships that may not be directly observable. The second level links this latent structure to the observed data, allowing the dependence information inferred at the hidden layer to be propagated to the observation process. By modelling the latent and observed components separately, the proposed framework provides greater flexibility in accommodating complex dependence patterns and reduces the likelihood of drawing misleading conclusions arising from an incorrectly specified observation-level model.

In practice, the process is not observed continuously but at discrete points.
\begin{align*}
 y_{i,t}=Y_{t}(k_{i,t})+ \nu_{i,t},~ t = 1,\ldots,T
 \end{align*}
where 
%\begin{math}
$\mathcal{T}_{t} =\{ k_{1,t},\ldots, k_{m_{t},t}\}$
%\end{math}
are observation points of $Y_{t}$ and $\nu_{i,t}$ = $\nu_{t}(k_{i,t})$,
%\\
%\textcolor{red}{The proposition is very abrupt. It needs some discussion? What are u trying to say? You started with $Y_t$, so what is $X_t$ below? You have not mentioned the FARMA process before, should you say something before you give this proposition? Dont use the "'//" everywhere. This creates a new line. Use can just use $a$ instead of beginning and ending math. For formulas like models, best to use align or align*. I have fixed these below, for you to use as examples.} \textcolor{blue}{(resolved)}
%
Thus, the two level hierarchical model, denoted by Model (\ref{initModel}), is given as: %\\
\begin{equation}
\label{initModel}
\begin{split}
 y_{i,t} & =\mu(k_{i,t})+\alpha_{t}(k_{i,t})+ \nu_{i,t}, \\
 \alpha_{t}(u) & =\int_{0}^{1} \psi (u,s) \alpha_{t-1}(s)\,ds + \epsilon_{t}(u) \ , \forall \ u \in \mathcal{T},
 \end{split}
\end{equation} %\\
where we assume that 
%\begin{math}
$\{\nu_{i,t}\}$ %\end{math} 
and 
%\begin{math} 
$\{\epsilon_{t}\}$ %\end{math} 
are mutually independent sequences.
Assume, $ \int_{0}^{1} \int_{0}^{1} \psi^2(k, s) \,dk \,ds < 1 $ for the stationarity of the $\{\alpha_t\}_{t=1}^{T}$ process.
For implementation of \eqref{initModel}, a finite set of evaluation points \begin{math} \mathcal{T}_{e} = \{k_{1},\ldots,k_{M}\} \in \mathcal{T} \end{math} %is to 
must be selected. Assume, 
%\begin{math}
$\mathcal{T}_{t} \subseteq \mathcal{T}_{e} \ \forall t$.
%\end{math}\ t. \\
Using quadrature methods, %one can 
it is possible to accurately approximate the integral in (\ref{initModel}) by choosing \begin{math} M \end{math} large and $\mathcal{T}_{e}$ to be dense in $\mathcal{T}$ such that
%\begin{math}
\begin{align*}
 \int_{0}^{1} \psi(k,s)\alpha(s)\,ds \approx (\psi(k,k_{1}),\ldots,\psi(k,k_{{M}})) \mathbf{Q}\boldsymbol{\alpha},
 \end{align*}
%\end{math} 
\\
where $\mathbf{Q}$ is a known quadrature weight matrix and 
%\begin{math} 
$\boldsymbol{\alpha} = (\alpha(k_{1}),\ldots,\alpha(k_{M}))'$. %\end{math}
%
%\textcolor{red}{Are u concatenating different vectors above? If yes, inside vectors must also be transposed, like I did?} \textcolor{blue}{resolved}
%

%\textcolor{red}{There may be a clever way of numbering different parts of some equations so you do not need to rewrite many eqns, especially in (7)-(9)? }\textcolor{blue}{resolved}

Let $\mathbf{Z}_t$ be the $m_{t}\times M$ incidence matrix that identifies the observation points at time $t$. Then, a state space representation of Model (\ref{initModel}) as a dynamic linear model (\cite{west2006bayesian}) in $\mathbf{\alpha}_{t}$, denoted by {Model (\ref{ModelEq}}) is as follows:

%\\

\begin{subequations}\label{ModelEq}
\begin{align}
 \mathbf{y}_{t} &= \mathbf{Z}_t \boldsymbol{\mu} + \mathbf{Z}_t \boldsymbol{\alpha}_{t} + \boldsymbol{\nu}_{t}, \quad && t=1, 2, \ldots, T \label{ModelEq_y} \\
 \boldsymbol{\nu}_{t} &\stackrel{\text{indep}}{\sim} \mathcal{N}(\mathbf{0}, \sigma_\nu^{2}\mathbf{I}_{m_t}), && t=1, 2, \ldots, T \label{noise} \\
 \boldsymbol{\alpha}_{t} &= \mathbf{\Psi} \mathbf{Q}\boldsymbol{\alpha}_{t-1} + \boldsymbol{\epsilon}_{t}, \quad \boldsymbol{\epsilon}_{t} \stackrel{\text{indep}}{\sim} \mathcal{N}(\mathbf{0}, \mathbf{K}_\epsilon), && t=2, \ldots, T \label{ModelEq_alpha_t} \\
 \boldsymbol{\alpha}_{1} &\sim \mathcal{N}(\mathbf{0}, \mathbf{K}_\epsilon) \nonumber
\end{align}
\end{subequations}

where 
\begin{align*}
 \mathbf{y}_{t} &= (y_{1,t}, \dots, y_{m_{t},t})', 
 & \boldsymbol{\mu} &= \big(\mu(k_1), \dots, \mu(k_M)\big)', \\
 \mathbf{\Psi} &= \big\{ \psi(k_{i},k_{j}) \big\}_{i,j=1}^{M}, 
 & \mathbf{K}_{\boldsymbol{\epsilon}} &= \big\{ \kappa_{\epsilon}(k_{i},k_{j}) \big\}_{i,j=1}^{M},
\end{align*}
and the quadrature weight matrix is defined as
\[
\mathbf{Q} = \text{diag}(w_1, \dots, w_M), \quad \text{where } 
w_k = \begin{cases} 
 \frac{h}{2}, & k = 1, M \\ 
 h, & k = 2, \dots, M-1, 
\end{cases} \quad \text{with } h = \frac{1}{M-1}.
\] %\\

%\textcolor{red}{Below, refer to $\sigma^2_\nu$, capturing the measurement error variance; not $\sigma_\nu$?} \textcolor{blue}{resolved}

In subsequent sections of this article, we % shall 
systematically address the potential distortions and structural anomalies associated with the key parameters of the model. Specifically, we focus our investigation on the mean vector $\boldsymbol{\mu}$, which governs the deterministic trend component; the scale parameter $\sigma_\nu^2$, capturing the measurement error variance;
and the transition matrix $\mathbf{\Psi}$, which dictates the dynamic stability and cross-dependencies within the latent state process.
%\\
%\textcolor{red}{After (2), maybe Write down in words about mu, sigmasqv, and Psi, because you need them again in Sec 2.2? }
%\textcolor{blue}{(resolved)}\\
%\textcolor{red}{Also, I think you shd explain Q a little more?}
%\textcolor{blue}{(resolved)}\\

\subsection{\MakeUppercase{A Unified Statistical Framework for Breaks in Parameters}} 

%\textcolor{red}{Note: I used eqref below to show equation numbers correctly. Also, why are you bounding equations? I dont see CJS allowing this? Best to do things their way. Can you redo the rest of the paper follwing my suggestions here and above? Then let us know, so we can continue to read?} \textcolor{blue}{(resolved)}

%\textcolor{red}{I wonder whether most of the first two paras belong in the Intro, and some parts to the main part of section 2. Here, start with para 3?}\textcolor{blue}{resolved}

To capture potential structural breaks across all facets of the process simultaneously, we propose a flexible framework that accounts for distinct change points in the (i) mean level, (ii) observation noise variance, and (iii) dynamic temporal dependencies. Specifically, let $\tau_1$, $\tau_2$, and $\tau_3$ denote the change points corresponding to the baseline mean vector, the observation error variance, and the autoregressive transition operator, respectively. The general structural break model, denoted by Model ~\eqref{Model7} is formulated as follows:

\begin{subequations}\label{Model7}
\begin{align}
 \mathbf{y}_t &= \begin{cases} 
 \mathbf{Z}_t \boldsymbol{\mu}_{1} + \mathbf{Z}_{t}\boldsymbol{\alpha}_{t} + \boldsymbol{\nu}_{t}, & t = 1, \dots, \tau_1 \\
 \mathbf{Z}_{t}\boldsymbol{\mu}_{2} + \mathbf{Z}_{t}\boldsymbol{\alpha}_{t} + \boldsymbol{\nu}_{t}, & t = \tau_1 + 1, \dots, T
 \end{cases} \label{Model7_obs} \\[10pt]
 \boldsymbol{\nu}_{t} &\stackrel{indep}{\sim} \begin{cases} 
 N(\boldsymbol{0}, \sigma_1^{2}\mathbf{I}_{m_t}), & t = 1, \dots, \tau_2 \\
 N(\boldsymbol{0}, \sigma_2^{2}\mathbf{I}_{m_t}), & t = \tau_2 + 1, \dots, T
 \end{cases} \label{Model7_noise} \\[10pt]
 \boldsymbol{\alpha}_t &= \begin{cases}
 \boldsymbol{\Psi}_{1} \mathbf{Q}\boldsymbol{\alpha}_{t-1}+\boldsymbol{\epsilon}_t, & t=2, \dots, \tau_3 \\
 \boldsymbol{\Psi}_{2} \mathbf{Q}\boldsymbol{\alpha}_{t-1}+\boldsymbol{\epsilon}_t, & t=\tau_3+1, \dots, T
 \end{cases}; \quad \boldsymbol{\epsilon}_{t} \stackrel{indep}{\sim} N(\boldsymbol{0},\mathbf{K}_\epsilon) \label{Model7_state} \\[10pt]
 \boldsymbol{\alpha}_{1} &\sim N(\boldsymbol{0}, \mathbf{K}_\epsilon), \nonumber
\end{align}
\end{subequations}
where the observation vector $\mathbf{y}_{t}$ and the state error covariance matrix $\mathbf{K}_{\boldsymbol{\epsilon}}$ are as in Model~\eqref{ModelEq}. Here, the baseline observation equation ~\eqref{ModelEq_y} and ~\eqref{noise} undergoes a structural decomposition governed by the distinct structural change points $\tau_1$ and $\tau_2$, which delineate shifts in the mean profile and measurement error variance, respectively. Similarly, the latent state transition dynamic~\eqref{ModelEq_alpha_t} is piecewise partitioned by the introduction of the regime shift parameter $\tau_3$, capturing variations in the state-evolution structure. The parameters $\boldsymbol{\mu}_1$ ($\boldsymbol{\mu}_2$), $\sigma_1^2$ ($\sigma_2^2$) and $\mathbf{\Psi}_1$ ($\mathbf{\Psi}_2$) represent the pre-change (post-change) baseline mean vector, observation error variance and autoregressive transition operator, respectively.

It is also possible that even when no explicit structural breaks occurred at the observational level, an underlying shift may occur within the unobserved state variables at time $\tau_3$. While a substantial body of work has focused on identifying structural changes in the observed functional data, such as changes in the mean function, observed variance, the considerably more challenging problem of detecting changes in the unobservable latent process remains unexplored. The primary difficulty arises from the fact that the latent states are not directly observable and must be inferred through a state-space framework. Consequently, any structural change in the latent dynamics is obscured by both measurement error and process noise, making the identification of change points substantially more difficult than in the observable setting. By modelling the dynamic evolution of the latent state process $\boldsymbol{\alpha}_t$, our framework can effectively capture and estimate these hidden structural changes. A novel approach that cannot be achieved by directly analysing observational data alone. 

%\textcolor{red}{I wonder whether you should proceed directly to Bayesian analysis after defining (3). Then separate subsections for (4) only, (5) only, and (6) only? Will that read well?} \textcolor{blue}{Decided not to}

The AR-operator transition model, which assumes a structural break solely in the transition operator $\mathbf{\Psi}$, constitutes a special case of Model~\eqref{Model7} and is formulated by coupling the baseline observation equation~\eqref{ModelEq_y}, ~\eqref{noise} along with ~\eqref{Model7_state}; with the change at $\tau$.

A distortion may occur in only a single parameter component while the remaining parameters remain time-invariant across the entire observation period. In such scenarios, the general formulation in ~\eqref{Model7} simplifies 
to single change-point models, where structural shifts are isolated to either the baseline mean at $\tau_1$ or the error variance at $\tau_2$. Hence, the mean transition model is defined by the equations ~\eqref{Model7_obs} along with ~\eqref{noise} and ~\eqref{ModelEq_alpha_t}. Likewise, the error variance change point model is defined by ~\eqref{Model7_noise} along with ~\eqref{ModelEq_y} and ~\eqref{ModelEq_alpha_t}.

Likewise, structural breaks may occur simultaneously across two parameter components while the third parameter remains constant throughout the time series. These configurations give rise to dual change-point models that allow for simultaneous shifts across parameter pairs. Concurrent breaks can occur in the mean level and error variance ($\tau_1, \tau_2$), as specified by~\eqref{Model7_obs} and~\eqref{Model7_noise} alongside ~\eqref{ModelEq_alpha_t}. Similarly, simultaneous shifts can affect the mean level and latent dynamics ($\tau_1, \tau_3$) governed by~\eqref{Model7_obs}, \eqref{noise}, and~\eqref{Model7_state}. There can also be shifts in the error variance and latent dynamics ($\tau_2, \tau_3$), described by~\eqref{Model7_noise} and~\eqref{Model7_state} together with~\eqref{ModelEq_y}.

\section{\MakeUppercase{Bayesian Inference and Computation}} 

%\textcolor{red}{Everywhere you refer to $\sigma_v$ but you are in fact modeling the variance? Shd you fix? }\textcolor{blue}{resolved}

In Section 3.1, we describe prior assumptions on the model parameters. Section 3.2 describes the initialization procedure of the MCMC. Section 3.3 provides the full conditional distributions of the parameters under the different models.

\subsection{PRIOR DISTRIBUTIONS }{\label{Prior}}
The parameters under the modelling assumption of \eqref{ModelEq} are $\boldsymbol{\mu}$ , $\sigma_\nu^2$, $\boldsymbol{\Psi}$ and $\boldsymbol{K_\epsilon}$. The introduction of change points decomposes $\boldsymbol\mu$ into $\boldsymbol{\mu}_1$, $\boldsymbol{\mu}_2$; $\sigma_\nu^2$ into $\sigma_1^2$ and $\sigma_2^2$, $\boldsymbol{\Psi}$ into $\boldsymbol{{\Psi}_1}$ and $\boldsymbol{{\Psi}_2}$.We have used the same prior assumptions for all the parameters as in \cite{kowal2019functional}. In our setup, $\boldsymbol{\mu}_1 \overset{d}{=} \boldsymbol{\mu}_2 \overset{d}{=} \boldsymbol{\mu}$,
\begin{math}
 \sigma_1^{2} \overset{d}{=} \sigma_2^{2} \overset{d}{=} \sigma_\nu^{2}
\end{math} and \begin{math}
\boldsymbol{\Psi_1} \overset{d}{=} \boldsymbol{{\Psi}_2} \overset{d}{=} \boldsymbol{{\Psi}}
\end{math}.

 $1.$ Prior distribution for the mean function $\boldsymbol{\mu}$ :

 \[
 \mu(k) = \boldsymbol{b^T(k)\theta_{\mu}} \; \forall \ k \in \mathcal{T}
\]\\
where, \begin{math}
 \boldsymbol{b^T} 
\end{math}
is the low rank thin plate spline basis and 
\[
 \boldsymbol{\theta_{\mu}} \sim N(\boldsymbol 0,\boldsymbol\Lambda_{\mu})\\ \quad \
 \text{with} \; \boldsymbol{\Lambda_{\mu}} = \text{Diag}(10^8,10^8,\lambda_{\mu}^{-1},\ldots,\lambda_{\mu}^{-1})\\ 
 \]
 and \[
 \lambda_{\mu}^{-1/2} \sim U(0,10^4).
\] 

$2.$ \[
\sigma_{\nu}^{2} \sim \text{Inverse Gamma}(10^{-3},10^{-3}). 
\]

%\textcolor{red}{Are s and u bold or scalar? They look diff. in diff. places.}
%\textcolor{blue}{resolved}

$3.$
\[
\psi(s,u)=\boldsymbol{(b_{\psi}}^{T}(u) \otimes \boldsymbol{b_{\psi}}^{T}(s)) \boldsymbol{\theta_{\psi}} 
\quad 
\text{for 
s , u} \in \mathcal{T}
\]\\
where
\begin{math} \boldsymbol{b_{\psi}^{T}} \end{math}is the cubic B-spline basis and 
\[
\boldsymbol{\theta_{\psi} \sim N(0,\tilde{\lambda}_{\psi}^{-1}\Omega_{\psi}^{-1})}\\
\]
s.t.
\begin{math}
\lambda_{\psi}=\tilde{\zeta}^{-2}\tilde{\lambda}_{\psi}
\end{math} where,
\begin{math}
\tilde{\zeta} \sim N(0,10^6) 
\end{math}
and
\begin{math}
 \tilde{\lambda}_{\psi} \sim Gamma(1/2,1/2)
\end{math}.

\[
\boldsymbol{\Omega_{\psi}}
=
\boldsymbol{\Omega_{2}}
+
\kappa \quad \boldsymbol{\Omega_{0}},
\]
where
\[
\log(\kappa)\sim N(0,4),
\]
\[
\boldsymbol\Omega_{0}
=
\iint_{\mathcal{T}^{2}}
\psi^{2}(s,u)\,ds\,du,
\]
and
\[
\boldsymbol\Omega_{2}
=
\iint_{\mathcal{T}^{2}}
\left[
\left(\frac{\partial^{2}\psi(s,u)}{\partial s^{2}}\right)^{2}
+
2\left(\frac{\partial^{2}\psi(s,u)}{\partial s\,\partial u}\right)^{2}
+
\left(\frac{\partial^{2}\psi(s,u)}{\partial u^{2}}\right)^{2}
\right]
\,ds\,du.
\]

$4.$ Prior Distribution for evolution error, $\boldsymbol\epsilon_t$:\\
The evolution error is represented as a dynamic factor model:
\[
\epsilon_t(u)=\sum_{j=1}^{J_\epsilon} e_{j,t}\phi_j(u)+\eta_t(u),
\]
where $J_\epsilon$ denotes the number of factors, $\{\phi_j(u)\}_{j=1}^{J_\epsilon}$ are the factor loading curves, $e_{j,t}$ are the corresponding factor scores, and $\eta_t(u)$ is a mean-zero approximation error.
\\
\vspace{3mm}
$(a)$ Prior Distribution for $\boldsymbol{\eta_t}$:
\[
\boldsymbol{\eta_t} \stackrel{\text{ind}}{\sim} \mathcal{GP}(\boldsymbol0, \boldsymbol{K_\eta}),
\]

where the covariance kernel is given by

\[
K_\eta(u,s)=\sigma_\eta^{2}\,\mathds{1}(u=s),
\]

and the prior distribution for the precision parameter is

\[
\sigma_\eta^{-2}\sim \mathrm{Gamma}\left(10^{-3},\,10^{-3}\right).
\]\\
\vspace{3mm}
\noindent
(b) {Prior Distribution for $\boldsymbol {e_t}$:}

\[
\boldsymbol{e_t \sim N(0,\Sigma_e)},
\]

where

\[
\boldsymbol\Sigma_e=\operatorname{diag}\left(\sigma_1^2,\sigma_2^2,\ldots,\sigma_{J_\epsilon}^2\right).
\]

The prior for the last precision parameter is given by

\[
\sigma_{J_\epsilon}^{-2}
\sim \mathrm{Gamma}\left(10^{-3},\,10^{-3}\right),
\]

and, for \(j=1,\ldots,J_\epsilon-1\),

\[
\left(\sigma_j^{-2}\mid\sigma_{j+1}^{-2}\right)
\sim
\mathrm{Uniform}\!\left(0,\sigma_{j+1}^{-2}\right).
\]
{(c) Prior Distribution for $\boldsymbol {\phi_j}$, $j=1,\ldots,J_\epsilon$:}

\[
\phi_j(u)=\mathbf{{b}^{\top}}(u)\boldsymbol{{\theta}_{\phi}},
\]

where $\mathbf{b}(u)$ denotes a spline basis function and

\[
\boldsymbol{\theta_{\phi}\sim N\left(\mathbf{0},\Lambda_{\phi}\right)},
\]

with

\[
\boldsymbol\Lambda_{\phi}
=
\operatorname{Diag}
\left(
10^{8},
10^{8},
\lambda_{\phi}^{-1},
\lambda_{\phi}^{-1},
\ldots,
\lambda_{\phi}^{-1}
\right).
\]

Finally,

\[
\lambda_{\phi}^{-1/2}\sim U(0,10^{4}).
\]

$5.$ Let $\tau$ denote a generic change-point location. For models involving simultaneous change-points, the notation $\tau$ is used generically to represent any of the change-point locations $\tau_1$, $\tau_2$, or $\tau_3$.

 Prior Distribution of 
\begin{math}
\tau:
\end{math}
\begin{align*}
\pi(\tau) &= \frac{1}{T-2} \quad \text{for } \tau \in \{2, 3, \ldots, T-1\}
\end{align*}
There can be several other choices for the prior distribution of the change-point location $\tau$, including geometric, negative binomial, and other informative priors that incorporate prior knowledge regarding the likely occurrence of a change. In this work, however, we adopt a discrete uniform prior for $\tau$, following \cite{li2021bayesian}. The discrete uniform prior is one of the most commonly used non-informative priors in Bayesian change-point analysis because it assigns equal probability to every admissible change-point location, thereby avoiding any undue preference for a particular time point. This choice is particularly appropriate when no reliable prior information regarding the location of the change point is available. Moreover, it provides a simple and computationally convenient formulation while allowing the posterior distribution to be driven primarily by the observed data rather than subjective prior beliefs.

%\textcolor{red}{Do we need Sec 3.2? RKF is well known - maybe move to appendix for now?}\textcolor{blue}{resolved}

\vspace{4pt}

\subsection{INITIALIZATION}
\label{initialization}
\textbf{Step 1:} Initialize $\tau_1,\tau_2 \quad \text{and} \quad  \tau_3$. \\
\textbf{Step 2:} Then estimate $\boldsymbol{{\mu}_1}$ as $\boldsymbol B(\sum_{i=1}^{\tau_1}\boldsymbol{B_{i}^{T}B_{i}+\Lambda_0})^{-1}(\sum_{i=1}^{\tau_1}\boldsymbol{B_iy_{i}^{*}})$ where $\boldsymbol{B}$ is the matrix of low rank thin plate spline basis functions, $\boldsymbol{B_{i}}$ is a submatrix of B having the rows equal to the indices where $\boldsymbol{y_{i}}$ has non missing observations. $\boldsymbol{y_{i}^{*}}$ denotes the column vector of $\boldsymbol{y_{i}}$ having only the non-missing observations. Similarly, $\boldsymbol{{\mu}_2}$ as $\boldsymbol B(\sum_{i=\tau_1+1}^{T}\boldsymbol{B_{i}^{T}B_{i}+\Lambda_0)}^{-1}(\sum_{i=\tau_1+1}^{T}\boldsymbol{B_iy_{i}^{*}})$ where $\boldsymbol \Lambda_0$ $= Diag(10^{-8},10^{-8},1,\ldots,1)$\\
\textbf{Step3 :}
A smoothing spline is first fitted to each centered observation $(\boldsymbol{y}_t-\boldsymbol{\mu})$, and the corresponding degrees of freedom are extracted. The median of these degrees of freedom is then used to fit smoothing splines to each $(\boldsymbol{y}_t-\boldsymbol{\mu})$, yielding the estimates of $\{\boldsymbol{\alpha}_t\}_{t=1}^{T}$. \\
\textbf{Step 4:} Estimate $\sigma_1^{2}$  and $\sigma_2^{2}$ based on $\tau_2$ given by \[
\sigma_1^{2}=
\frac{
\displaystyle \sum_{t=1}^{\tau_2}\sum_{i=1}^{m_t} 
\left(y_{it}-\mu(k_{it})-\alpha_t(k_{it})\right)^2
}{
\displaystyle \sum_{t=1}^{\tau_2} m_t
}
\quad \text{and} \ 
\sigma_2^{2}=  \frac{
\displaystyle \sum_{t=\tau_2+1}^{T}\sum_{i=1}^{m_t} 
\left(y_{it}-\mu(k_{it})-\alpha_t(k_{it})\right)^2
}{
\displaystyle \sum_{t=\tau_2+1}^{T} m_t
}
\]
\\
\textbf{Step 5:} $\boldsymbol{\Psi_1}$ is based on the parametrization 
\begin{math}
\boldsymbol{\Psi_1}=\boldsymbol{B_{\Psi_1}}\boldsymbol{\Theta_{\Psi_1}}\boldsymbol{B_{\Psi_1}^{'}}
\end{math} where $\boldsymbol{B_{\Psi_1}}$ is a matrix of cubic B-spline basis functions and
\begin{math}
\boldsymbol{\theta_{\Psi_1}} = vec(\boldsymbol{\Theta_{\Psi_1}}) 
\end{math} is sampled from N
\begin{math}
(\boldsymbol{A_{\psi_1}}\boldsymbol{a_{\psi_1}},\boldsymbol{A_{\psi_1}})
\end{math}
where,
\[
\boldsymbol{A_{\psi_1}^{-1}}=\boldsymbol{\Omega_{\psi_1}}+\boldsymbol{[(B_{\psi_1}^{'}Q)\{\sum_{t=2}^{\tau_3}{\alpha}_{t-1}{\alpha}_{t-1}^{'}\}(B_{\psi_1}^{'}Q)^{'}]\otimes[(B_{\psi_1}^{'}B_{\psi_1})]} \]
\[
\boldsymbol{a_{\psi_1}}=vec(\boldsymbol{B_{\psi_1}^{'}\sum_{t=2}^{\tau_3}{\alpha_{t}}{\alpha}_{t-1}^{'}(B_{\psi_1}^{'}Q)^{'}})
\] 
\textbf{Step 6:} $\boldsymbol{\Psi_2}$ is based on the parametrization 
\begin{math}
\boldsymbol{\Psi_2}=\boldsymbol{B_{\Psi_2}\Theta_{\Psi_2}B_{\Psi_2}^{'}}
\end{math} where $\boldsymbol{B_{\Psi_2}}$ is a matrix of cubic B-spline basis functions and
\begin{math}
\boldsymbol{\theta_{\Psi_2}} = vec(\boldsymbol{\Theta_{\Psi_2}}) 
\end{math} is sampled from N
\begin{math}
(\boldsymbol{A_{\psi_2}a_{\psi_2}},\boldsymbol{A_{\psi_2}})
\end{math}
where, 
\[
\boldsymbol{A_{\psi_2}^{-1}}=\boldsymbol{\Omega_{\psi_2}}+\boldsymbol{[(B_{\psi_2}^{'}Q)\{\sum_{t=\tau_3+1}^{T}{\alpha}_{t-1}{\alpha}_{t-1}^{'}\}(B_{\psi_2}^{'}Q)^{'}]\otimes[(B_{\psi_2}^{'}B_{\psi_2})]} \]
\[\boldsymbol{a_{\psi_2}}=vec(\boldsymbol{B_{\psi_2}^{'}\sum_{t=\tau_3+1}^{T}{\alpha_{t}}{\alpha}_{t-1}^{'}(B_{\psi_2}^{'}Q)^{'}})
\]
\textbf{Step 7:} Using the estimated FAR kernel, we compute the innovation functions, $\{\boldsymbol{\epsilon_{t}}\}_{t=1}^{T}$. The innovations are then decomposed into factor loading curves (FLCs), $\boldsymbol{\Phi}=(\phi_1,\ldots,\phi_{J_\epsilon})$, and time-dependent factor scores, ${\boldsymbol{e_t}}=(e_{1,t},\ldots,e_{J_\epsilon,t})^\top$. For simplicity, we assume that
\[
\boldsymbol{e_t \mid \Sigma_e} \stackrel{\mathrm{ind}}{\sim} N(\boldsymbol0,\boldsymbol{\Sigma_e}),
\]
where
\[
\boldsymbol{\Sigma_e}=\operatorname{diag}\left(\sigma_1^2,\ldots,\sigma_{J_\epsilon}^2\right),
\].

The FDLM parameters are initialized using the singular value decomposition (SVD) of the innovation matrix,
\[
(\epsilon_1,\ldots,\epsilon_T)^\top
=
\boldsymbol{U_eD_eV_e^\top}.
\]
The initial estimate of the factor loading matrix is obtained by setting
\[
\boldsymbol\Phi=\left(v_1,\ldots,v_{J_\epsilon}\right),
\]
where $v_j$ denotes the $j$th column of $V_e$. The corresponding initial factor scores are computed as
\[
\boldsymbol{E}=\boldsymbol{U_eD_e}(:,1:J_\epsilon),
\]
where $\boldsymbol{E}=(e_1,\ldots,e_T)^\top$.
Further, the variance components ${\{\sigma_j^2\}_{j=1}^{J_\epsilon}}$ and $\sigma_\eta^2$ are estimated using their conditional maximum likelihood estimators. The approximation error process $\boldsymbol {\{\eta_t\}}_{t=1}^{T}$ is assumed to be mean-zero Gaussian white noise over the functional domain, i.e.,
\[
\boldsymbol{\eta_t \stackrel{\mathrm{iid}}{\sim} \mathcal{GP}(0,K_\eta)},
\]
where
\[
K_\eta(\tau,u)=\sigma_\eta^2\,\mathds{1}(\tau=u).
\]

\textbf{Step 8:} The implied covariance matrix for $\boldsymbol{\eta}_t$ is $\mathbf{K}_\eta = \boldsymbol{\Phi} \boldsymbol{\Sigma}_e \boldsymbol{\Phi^\top} + \sigma_\eta^2 \mathbf{I}_M$, conditional on $\boldsymbol{\phi}_j$, $\sigma_j^2$, and $\sigma_\eta^2$.

\subsection{FULL CONDITIONAL DISTRIBUTIONS FOR GIBBS SAMPLING}{\label{posterior}}
In this section we %shall 
provide the full conditional distributions of the parameters under Model (\ref{Model7}). Steps~1--3 follow the methodology of \cite{kowal2019functional}, with appropriate modifications to accommodate the presence of a change point. Step~4 remains identical to that proposed in \cite{kowal2019functional}. 

\begin{enumerate}[label=\arabic*.]
\item 
\begin{enumerate}[label=(\alph*)]
\item
Sample 
\begin{math}
\{\boldsymbol{\alpha_{t}}\}_{t=1}^{T}|....
\end{math}
using Robust Kalman Filter of \cite{ruckdeschel2014robust}. 
\item The full conditional distributions for the two mean functions $\boldsymbol{\mu_1}$ and $\boldsymbol{\mu_2}$ are provided as follows.
\begin{align*}
 \mu_i(k) &= \mathbf{b^{\top}}(k)\boldsymbol{\theta}_{\mu_i},
\qquad \text{for } i=1,2.\\
 \intertext{$\left[\boldsymbol{\theta}_{\mu_1} \mid \dots\right] \sim \mathcal{N}(\mathbf{A}_{\mu_1}\mathbf{a}_{\mu_1}, \mathbf{A}_{\mu_1})$, where for $\tau_1 > \tau_2$:}
 \boldsymbol{\mathbf{A}_{\mu_1}^{-1}} &= \boldsymbol{\Lambda_{\mu_1}^{-1}} + \sigma_{1}^{-2} \boldsymbol{\sum_{t=1}^{\tau_2} B'Z_{t}'Z_{t}B} + \sigma_{2}^{-2} \boldsymbol{\sum_{t=\tau_2+1}^{\tau_1} B'Z_{t}'Z_{t}B} \\
 \mathbf{a}_{\mu_1} &= \sigma_{1}^{-2} \boldsymbol{\sum_{t=1}^{\tau_2} B'Z_{t}'(y_{t}-Z_{t}\alpha_{t})} + \sigma_{2}^{-2} \boldsymbol{\sum_{t=\tau_2+1}^{\tau_1} B'Z_{t}'(y_{t}-Z_{t}\alpha_{t})} \\
 \intertext{and for $\tau_1 \le \tau_2$:}
 \mathbf{A}_{\mu_1}^{-1} &= \boldsymbol{\Lambda_{\mu_1}^{-1}} + \sigma_{1}^{-2} \boldsymbol{\sum_{t=1}^{\tau_1} B'Z_{t}'Z_{t}B} \\
 \mathbf{{a}_{\mu_1}} &= \sigma_{1}^{-2} \boldsymbol{\sum_{t=1}^{\tau_1} B'Z_{t}'(y_{t}-Z_{t}\alpha_{t}}) \\[6pt]
\intertext{$\left[\boldsymbol{\theta}_{\mu_2} \mid \dots\right] \sim \mathcal{N}(\mathbf{A}_{\mu_2}\mathbf{a}_{\mu_2}, \mathbf{A}_{\mu_2})$ where for $\tau_1 \geq \tau_2$} 
 \mathbf{A}_{\mu_2}^{-1} &= \boldsymbol{\Lambda_{\mu_2}^{-1}} + \sigma_{2}^{-2} \boldsymbol{\sum_{t=\tau_1+1}^{T} B'Z_{t}'Z_{t}B_{\phi}} \\
 \mathbf{a}_{\mu_2} &= \sigma_{2}^{-2} \boldsymbol{\sum_{t=\tau_1+1}^{T} B'Z_{t}'(y_{t}-Z_{t}\alpha_{t}}) \quad \ ; \ \\ \\ 
 \intertext{and for $\tau_1 < \tau_2$:}
 \mathbf{A}_{\mu_2}^{-1} &= \boldsymbol{\Lambda_{\mu_2}^{-1}} + \sigma_{1}^{-2} \boldsymbol{\sum_{t=\tau_1+1}^{\tau_2} B'Z_{t}'Z_{t}B} + \sigma_{2}^{-2} \boldsymbol{\sum_{t=\tau_2+1}^{T} B'Z_{t}'Z_{t}B} \\
 \mathbf{a}_{\mu_2} &= \sigma_{1}^{-2} \boldsymbol{\sum_{t=\tau_1+1}^{\tau_2} B'Z_{t}'(y_{t}-Z_{t}\alpha_{t}}) + \sigma_{2}^{-2} \boldsymbol{\sum_{t=\tau_2+1}^{T} B'Z_{t}'(y_{t}-Z_{t}\alpha_{t}}) \\
 \intertext{where}
 \boldsymbol \Lambda_{\mu_i} &= \text{Diag}(10^{8}, 10^{8}, \lambda_{\mu_i}^{-1}, \dots, \lambda_{\mu_i}^{-1})
 \intertext{where $\boldsymbol b(k)$ is the basis function at $k$ and $\boldsymbol B = \boldsymbol {\left(b(\tau_1), \dots, b(\tau_M)\right)^\top}$ is the matrix of low rank thin plate spline basis functions. The smoothing parameter is sampled as:}
 \left[\lambda_{\mu_i} \mid \dots\right] &\sim \text{Gamma}\left(\frac{1}{2}(J_{\mu_i}-3), \; \frac{1}{2}\sum_{j=3}^{J_{\mu_i}}\theta_{\mu_i,j}^{2}\right)
 \intertext{where $J_{\mu_i}$ is the dimension of $\boldsymbol{{\theta}_{\mu_{i}}}$, and $\theta_{\mu_i,j}$ is its $j$-th component.}
\end{align*}
\end{enumerate}
\end{enumerate}

$2)$
Measurement error precision,
\begin{align*}
\left[\sigma_{1}^{-2} \mid \dots\right] &\sim \text{Gamma}\Bigg(10^{-3} + \frac{1}{2}\sum_{t=1}^{\tau_2} m_{t}, \; 10^{-3} + \frac{1}{2}\sum_{t=1}^{\tau_1}\sum_{i=1}^{m_{t}}\left(y_{i,t} - \mu_1(k_{i,t}) - \alpha_{t}(k_{i,t})\right)^{2}\Bigg) \quad \ ; \ \tau_1 \geq \tau_2
\\
\left[\sigma_{1}^{-2} \mid \dots\right] &\sim \text{Gamma}\Bigg(10^{-3} + \frac{1}{2}\sum_{t=1}^{\tau_2} m_{t}, \; 10^{-3} + \frac{1}{2}\sum_{t=1}^{\tau_1}\sum_{i=1}^{m_{t}}\left(y_{i,t} - \mu_1(k_{i,t}) - \alpha_{t}(k_{i,t})\right)^{2} \\
&\qquad\qquad + \frac{1}{2}\sum_{t=\tau_1 + 1}^{\tau_2}\sum_{i=1}^{m_{t}}\left(y_{i,t} - \mu_2(k_{i,t}) - \alpha_{t}(k_{i,t})\right)^{2}\Bigg) \quad \ ; \ \tau_1 < \tau_2 \\
\left[\sigma_{2}^{-2} \mid \dots\right] &\sim \text{Gamma}\Bigg(10^{-3} + \frac{1}{2}\sum_{t=\tau_2 + 1}^{T} m_{t}, \; 10^{-3} + \frac{1}{2}\sum_{t=\tau_2 + 1}^{\tau_1}\sum_{i=1}^{m_{t}}\left(y_{i,t} - \mu_1(k_{i,t}) - \alpha_{t}(k_{i,t})\right)^{2} \\
&\qquad\qquad + \frac{1}{2}\sum_{t=\tau_1 + 1}^{T}\sum_{i=1}^{m_{t}}\left(y_{i,t} - \mu_2(k_{i,t}) - \alpha_{t}(k_{i,t})\right)^{2}\Bigg) \quad \ ; \ \tau_1 > \tau_2 \\
\left[\sigma_{2}^{-2} \mid \dots\right] &\sim \text{Gamma}\Bigg(10^{-3} + \frac{1}{2}\sum_{t=\tau_2 + 1}^{T} m_{t}, \; 10^{-3} + \frac{1}{2}\sum_{t=\tau_2 + 1}^{T}\sum_{i=1}^{m_{t}}\left(y_{i,t} - \mu_2(k_{i,t}) - \alpha_{t}(k_{i,t})\right)^{2}\Bigg) 
\quad \ ; \ \tau_1 \leq \tau_2 
\end{align*}

 3. The FAR kernel $\psi$: 
 The kernel is defined by tensor product of basis functions :
 \begin{align*}
 \psi_{i}(s,u) &=\boldsymbol{ b_{\psi_{i}}(s)^{\prime}\Theta_{\psi_{i}}b_{\psi_{i}}(u) = (b_{\psi_{i}}^{T}(u) \otimes b_{\psi_{i}}^{T}(s))\theta_{\psi_{i}} } \quad ;\ \text{for i=1,2}
 \end{align*} where
 $\boldsymbol{\theta_{\psi_{i}}}$ = $\tilde{\xi}_{\psi_{i}}\boldsymbol{\tilde{\theta}_{\psi_{i}}}$ and
$\boldsymbol{\theta_{\psi_i}}$=
vec($\boldsymbol{\Theta_{\psi_i}}$). 
 Let, $\boldsymbol{B_{\psi_{i}}$ = $(b_{\psi_{i}}(k_{1})$, \ldots, $b_{\psi_{i}}(k_{M}))^{\prime}}$ where $\boldsymbol {b_{\psi_{i}}(k)}$ is the basis function at k. 

(a) sample 
\begin{math}
[\boldsymbol{\tilde{\theta}_{\psi_{i}}^{'}}|...]\sim
\end{math}
N
\begin{math}
(\boldsymbol{A_{\psi_{i}}a_{\psi_{i}},A_{\psi_{i}}}),
\end{math}
where 
\[
\boldsymbol{A_{\psi_{i}}^{-1}}=\tilde{\lambda}_{\psi_{i}}\boldsymbol\Omega_{\psi_{i}}+\xi_{\psi_{i}}^{2}\boldsymbol{[(B_{\psi_{i}}^{'}Q)\{\sum_{t=t_{1i}}^{t_{2i}}\alpha_{t-1}\alpha_{t-1}^{'}\}(B_{\psi_{i}}^{'}Q)^{'}]\otimes[(B_{\psi_{i}}^{'}K_{\epsilon}^{-1}B_{\psi_{i}})]}
\]
\[
\boldsymbol{a_{\psi_{i}}}=\tilde{\xi}_{\psi_{i}}vec(\boldsymbol{B_{\psi_{i}}^{'}K_{\epsilon}^{-1}\sum_{t=t_{1i}}^{t_{1i}}\alpha_{t}\alpha_{t-1}^{'}(B_{\psi_{i}}^{'}Q)^{'})}
\]
(b)
sample 
\begin{math}
 [\tilde{\xi}_{\psi_{i}}|...]\sim
\end{math}
N
\begin{math}
(A_{\tilde{\xi}_{\psi_{i}}}a_{\tilde{\xi}_{\psi_{i}}},A_{\tilde{\xi}_{\psi_{i}}})
\end{math}, where 
\[
A_{\tilde{\xi}_{\psi_{i}}}^{-1}
=10^{-6}+\boldsymbol{\tilde{\theta}_{\psi_{i}}^{'}[(B_{\psi_{i}}^{'}Q)\{\sum_{t=t_{1i}}^{t_{2i}}\alpha_{t-1}\alpha_{t-1}^{'}\}(B_{\psi_{i}}^{'}Q)^{'}]\otimes[(B_{\psi_{i}}^{'}K_{\epsilon}^{-1}B_{\psi_{i}})]\tilde{\theta}_{\psi_{i}}}
\]
\[
a_{\tilde{\xi}_{\psi_{i}}}=\boldsymbol{\tilde{\theta}_{\psi_{i}}}^{'}vec(\boldsymbol{B_{\psi_{i}}^{'}K_{\epsilon}^{-1}\sum_{t=2}^{\tau}\alpha_{t}\alpha_{t-1}^{'}(B_{\psi_{i}}^{'}Q)^{'}})
\]
\begin{align*}
 \intertext{(c) Sample} \tilde{\lambda}_{\psi_i} \sim \text{Gamma} \left(\frac{1}{2} + \frac{J_{\psi_i}^{2}}{2}, \frac{1}{2} + \frac{\boldsymbol{\theta_{\psi_i}' \Omega_{\psi_i} \theta_{\psi_i}}}{2}\right).
 \intertext{where $J_{\psi_i}$ is the dimension of $\boldsymbol{\theta}_{\psi_i}$.}
\end{align*}
 where
 \begin{equation*}
t_{1i} = 
\begin{cases} 
 2 & \text{for } i=1, \\ 
 \tau_3+1 & \text{for } i=2. 
\end{cases}
\end{equation*} and 
\begin{equation*}
t_{2i} = 
\begin{cases} 
 \tau_3 & \text{for } i=1, \\ 
 T & \text{for } i=2. 
\end{cases}
\end{equation*} 

$4.$ 
The innovation covariance, $K_{\epsilon}$, under the FDLM:

\begin{align*}
 \intertext{ (a) The factors $\boldsymbol {\{e_t\}_{t=1}^{T}}$ using the prior $\boldsymbol{e_{t}\sim N(0,\Sigma_{e})}$ and the conditional likelihood}
 \boldsymbol{ \epsilon_{t}} &= \boldsymbol{\alpha_{t}-\Psi Q \alpha_{t-1}=\Phi e_{t}+\eta_{t}}, \\
 \intertext{We sample $\boldsymbol{\left[e_{t}|\dots\right]} \sim N(\boldsymbol{A_{e}a_{e_{t}},A_{e}})$, where:}
 \boldsymbol{A_{e}^{-1}} &= \sigma_{\eta}^{-2}\boldsymbol{\Phi^{\prime}\Phi+\Sigma_{e}^{-1}} \\
 &= \text{Diag}\left(\left\{\sigma_{\eta}^{-2}+\sigma_{j}^{-2}\right\}_{j=1}^{J_{\epsilon}}\right) \\
 \boldsymbol{a_{e_{t}}} &= \sigma_{\eta}^{-2}\boldsymbol{\Phi^{\prime}\epsilon_{t}} \\
 \intertext{Since $\boldsymbol A_{e}$ is time-invariant and diagonal, we can sample $\boldsymbol{\{e_t\}_{t=1}^{T}}$ jointly and efficiently.}
\end{align*}

\begin{align*}
 \intertext{(b) The factor precisions, $\sigma_{j}^{2}$: First, sample the final precision:}
 \left[\sigma_{J_{\epsilon}}^{-2}\right] &\sim \text{Gamma}\left(10^{-3}+\frac{T}{2}, 10^{-3}+\frac{1}{2}\sum_{t=1}^{T}e_{J_{\epsilon},t}^{2}\right) \\
 \intertext{Then, for $j=J_{\epsilon}-1, \ldots, 1$, set:}
 \sigma_j^{-2} &= F_{\phi}^{-1}(U; s_\phi, r_{\phi_j}) \\
 \intertext{where $F_\phi$ is the distribution function for a Gamma random variable with parameters:}
 s_\phi &= \frac{(T-1)}{2} \\
 r_{\phi_j} &= \sum_{t=1}^T \frac{e^2_{j,t}}{2} \\
 \intertext{and $U$ is drawn from a truncated uniform distribution:}
 U &\sim \text{Uniform}(a_{\phi_j},b_{\phi_j}) \\
 \intertext{where the truncation bounds are defined by the previous precision $\sigma_{j+1}^{-2}$:}
 a_{\phi_j} &= F_\phi(0;s_\phi,r_{\phi_j}) \\
 b_{\phi_j} &= F_\phi(\sigma^{-2}_{j+1};s_\phi,r_{\phi_j})
\end{align*}

 \begin{align*}
 \intertext{(c) The approximation error precision, $\sigma_{\eta}^{-2}$ : sample}
 \left[\sigma_{\eta}^{-2}|\dots\right] &= \text{Gamma}\left(10^{-3}+\frac{TM}{2}, 10^{-3}+\frac{1}{2}\sum_{t=1}^{T}|| \boldsymbol{\epsilon_{t}-\Phi e_t}||^{2}\right)
\end{align*}

 \begin{align*}
 \intertext{(d) The factor loading curves: for $j=1, \ldots, J_{\epsilon}$ (in random order), sample $\boldsymbol{\xi_{j}}$ as:}
 \boldsymbol{ \xi_{j}} &\sim N(\boldsymbol{A_{\xi_{j}}a_{\xi_{j}},A_{\xi_{j}}}), \\
 \intertext{where the components are defined as:}
 \boldsymbol{A_{\xi_{j}}^{-1}} &= \boldsymbol{\Lambda_{\phi_{j}}^{-1}}+\sigma_{\eta}^{-2}\left(\sum_{t=1}^{T}e_{j,t}^{2}\right)\boldsymbol{B_{\phi}^{\prime}B_{\phi} }\\
 \boldsymbol{a_{\xi_{j}}} &= \sigma_{\eta}^{-2}\boldsymbol{B_{\phi}}^{\prime}\left(\sum_{t=1}^{T}e_{j,t}\right)\boldsymbol{\left(\epsilon_{t}-B_{\phi}\sum_{k\neq j}\xi_{k}e_{k,t}\right)} \\
 \text{where } \boldsymbol{B_{\phi}} \text{ is a matrix of low rank thin plate spline basis functions.}\\
 \intertext{To enforce the orthogonality constraint, we condition on the linear constraints}
 \boldsymbol{B_{\phi}\xi_{k}^{\prime}B_{\phi}\xi_{j}} &= \boldsymbol{0} \quad \text{for } k\neq j; \\
 \intertext{since $\boldsymbol{\xi_{j}}$ is Gaussian and $\boldsymbol{\xi_{k}}$ is conditioned upon, the resulting distribution is Gaussian with easily computable moments, which is also convenient for efficient sampling.}
 \intertext{The smoothing parameters are sampled:}
 \left[\lambda_{\phi_{j}}|\dots\right] &= \text{Gamma}\left(\frac{1}{2}(J_{\phi}-3),\frac{1}{2}\sum_{k=3}^{J_{\phi}} \xi_{j,k}^{2}\right) \\
 \intertext{where $\xi_{j,k}$ is the $k$-th component of $\xi_{j}$.}
\end{align*}
In case of only a baseline shift, we have $\sigma_1^{2}$ = $\sigma_2^{2}=\sigma_\nu^2$ and $\boldsymbol{\Psi_1=\Psi_2=\Psi}$. For the change in observational error variance, $\boldsymbol{\mu_1=\mu_2=\mu}$ and $\boldsymbol{\Psi_1=\Psi_2=\Psi}$ whereas for the operator transition, $\boldsymbol{\mu_1=\mu_2=\mu}$ and $\sigma_1^{2}=\sigma_2^{2}=\sigma_\nu^2$. In both cases, Kalman smoother has been used to estimate $\{\alpha_t\}_{t=1}^{T}$. Similarly, for two parameter changes, the third remains unsplit.

Having stated the full conditional distributions, we proceed to the methodologies of estimating the change points.

\section{ALGORITHMS}\label{algo}

We propose different algorithms for detecting change point locations. These methods leverage blocked Gibbs sampling techniques that explicitly incorporate the dependence inherent in the structured pattern. After initializing the parameter values according to ~\eqref{initialization}, we proceed to the MCMC step. At each iteration, we evaluate the full conditional probability distribution of the change point location and define the estimate $\tau$ to be the change point location that maximizes the probability. Here, we first present the general change point algorithm,  followed by single parameter change point algorithms for the mean and observational error variance. \\
%\textcolor{violet}{If you define $y-\mu_1-\alpha$ as $z_1^{(s)}(k_{i,t})$, $y-\mu_2-\alpha$ as $z_2$ and $y-\mu-\alpha$ as $z$ then D1, D2, D3 etc, as well as the expression in algorithm3 are simplified}

\hrule
\captionof{algorithm}{Iterative Gibbs Sampling for Simultaneous Change-point Detection in Mean -Volatility - AR operator (Model ~\eqref{Model7})}
\label{algo7}
\hrule
\begin{algorithmic}[1]

\STATE \textbf{Initialization:} 
 Set the initial values for $\tau_1^{(1)}, \tau_2^{(1)}, \tau_3^{(1)}$ and hence estimate  $\mu_1^{(1)}$, $\mu_2^{(1)}$, $\{{\alpha}_t^{(1)}\}$, $\sigma_1^{(1)}$, $\sigma_2^{(1)}$, $\Psi_1^{(1)}$, $\Psi_2^{(1)}$ and $K_{\epsilon}^{(1)}$ according to Section ~\eqref{initialization}. Further re-estimate $\{\boldsymbol{\alpha}_t^{(1)}\}_{t=1}^{T}$ using a \textbf{Robust Kalman Filter}. 

    \STATE \textbf{For} $s=1,\ldots,S$ \textbf{do}

\hspace{\algorithmicindent}\textbf{(a) Computing probability distribution of $\tau_3$:} For each candidate $j \in \{2, \dots, T-1\}$, compute :
    \begin{equation}{\label{eqn11}}
   p_{3}^{(s)}\!\left(j\right)=\frac{\frac{1}{|K_\epsilon^{(s)}|^{T/2}} e^{-\frac{1}{2}(\sum_{t=2}^{j} (\mathbf{\alpha}^{(s)}_t-\Psi_1^{(s)} Q \mathbf{\alpha}^{(s)}_{t-1})^{T}K_\epsilon^{{(s)}^{-1}}(\mathbf{\alpha}^{(s)}_t-\Psi_1^{(s)} Q \mathbf{\alpha}^{(s)}_{t-1})+\sum_{t=j+ 1}^T(\mathbf{\alpha}_t-\Psi_2^{(s)} Q \mathbf{\alpha}^{(s)}_{t-1})^{T}K_\epsilon^{{(s)}^{-1}}(\mathbf{\alpha}^{(s)}_t-\Psi_2^{(s)} Q \mathbf{\alpha}^{(s)}_{t-1}))}}{ \sum_{k=2}^{T-1}\frac{1}{|K_\epsilon^{(s)}|^{T/2}} e^{-\frac{1}{2}(\sum_{t=2}^{k}(\mathbf{\alpha}^{(s)}_t-\Psi_1^{(s)} Q \mathbf{\alpha}^{(s)}_{t-1})^{T}K_\epsilon^{{(s)}^{-1}}(\mathbf{\alpha}^{(s)}_t-\Psi_1^{(s)} Q \alpha^{(s)}_{t-1})+\sum_{t=k+ 1}^T(\mathbf{\alpha}^{(s)}_t-\Psi_2^{(s)} Q\mathbf{\alpha}^{(s)}_{t-1})^{T}K_\epsilon^{{(s)}^{-1}}(\mathbf{\alpha}^{(s)}_t-\Psi_2^{(s)} Q \mathbf{\alpha}^{(s)}_{t-1}))}}
  \end{equation}
    
   \hspace{\algorithmicindent}\textbf{(b) Coordinate Update:} Update the change-point $\tau_{3}^{(s+1)}$  by setting \[
\tau_{3}^{(s+1)}
=
\operatorname*{arg\,max}_{j \in \{2,\ldots,T-1\}}
p_{3}^{(s)}\!\left(j\right).
\]
    
 \hspace{\algorithmicindent}\textbf{(c)  Computing probability distribution of $\tau_2$}:
    
    \hspace{\algorithmicindent}\textbf{(i)} Compute
     \begin{equation}
p^{(s)}_{2}\!\left(j\right) =
\frac{
\displaystyle
\frac{1}{
\left (\sigma_1^{(s)}\right)^{\sum_{t=1}^{j} m_t}
\left(\sigma_2^{(s)}\right)^{\sum_{t=j+1}^{T} m_t}
}
\exp\Bigg[
-\frac{1}{2}
D_1^* \Bigg]}{
\displaystyle
\sum_{\substack{j=2 \\ j > \tau_1^{(s)}}}^{T-1}
\frac{1}{
\left (\sigma_1^{(s)}\right)^{\sum_{t=1}^{j} m_t}
\;\left(\sigma_2^{(s)}\right)^{\sum_{t=j+1}^{T} m_t}
}
\exp\Bigg[
-\frac{1}{2}
D_1^* \Bigg]
}
\end{equation}
Define, 
\[
\begin{gathered}
    z_{1,t}^{(s)}=\left(y_{i,t}-\mu_1^{(s)}(k_{i,t})-\alpha_t^{(s)}(k_{i,t})\right)^2 \\
    z_{2,t}^{(s)}= \left(y_{i,t}-\mu_2^{(s)}(k_{i,t})-\alpha_t^{(s)}(k_{i,t})\right)^2
\end{gathered}
\]

\[
\begin{gathered}
    D_1^*=\Bigg( \frac{ \sum_{t=1}^{\tau_1^{(s)}} \sum_{i=1}^{m_t} z_{1,t}^{(s)} + \sum_{t=\tau_1^{(s)}+1}^{j} \sum_{i=1}^{m_t} z_{2,t}^{(s)} }{ \sigma_1^{(s)2} } + \frac{ \sum_{t=j+1}^{T} \sum_{i=1}^{m_t} z_{1,t}^{(s)} }{ \sigma_2^{(s)2} } \Bigg)
\end{gathered}
\]

%\[
% \begin{gathered}
% D_1^*=\Bigg(
% \frac{
% \sum_{t=1}^{\tau_1^{(s)}} \sum_{i=1}^{m_t}
% \left(y_{i,t}-\mu_1^{(s)}(k_{i,t})-\alpha_t^{(s)}(k_{i,t})\right)^2
% +
% \sum_{t=\tau_1^{(s)}+1}^{j} \sum_{i=1}^{m_t}
% \left(y_{i,t}-\mu_2^{(s)}(k_{i,t})-\alpha_t^{(s)}(k_{i,t})\right)^2
% }{
% \sigma_1^{(s)2}
%}
% + \\
%\frac{
%\sum_{t=j+1}^{T} \sum_{i=1}^{m_t}
%\left(y_{i,t}-\mu_1^{(s)}(k_{i,t})-\alpha_t^{(s)}(k_{i,t})\right)^2
%}{
%\sigma_2^{(s)2}
%}
%\Bigg)
%\Bigg]
%\end{gathered}
%\]

for each $j$ 
\begin{math}
    \in \{\tau_1^{(s)}+1,\ldots, T-1\}
\end{math}.

\hspace{\algorithmicindent}\textbf{(ii)} 
     Compute 
\begin{equation}
p^{(s)}_{2}\!\left(j\right) =
\frac{
\displaystyle
\frac{1}{
\left (\sigma_1^{(s)}\right)^{\sum_{t=1}^{j} m_t}
\left(\sigma_2^{(s)}\right)^{\sum_{t=j+1}^{T} m_t}
}
\exp\Bigg[
-\frac{1}{2}
D_2^* \Bigg]}{
\displaystyle
\sum_{\substack{j=2 \\ j < \tau_1^{(s)}}}^{T-1}
\frac{1}{
\sigma_1^{(s)\,\sum_{t=1}^ {j} m_t}
\;\sigma_2^{(s)\,\sum_{t=j+1}^T m_t}
}
\exp\Bigg[
-\frac{1}{2}
D_2^* \Bigg]
}
\end{equation}
\[
\begin{gathered}
D_2^*=\Bigg(
\frac{
\sum_{t=1}^{j} \sum_{i=1}^{m_t} z_{1,t}^{(s)}
}{
\sigma_1^{(s)2}
}
+
\frac{
\sum_{t=j+1}^{\tau_1^{(s)}} \sum_{i=1}^{m_t} z_{1,t}^{(s)} + \sum_{t=\tau_{1}^{(s)}+1}^{T} \sum_{i=1}^{m_t} z_{2,t}^{(s)}
}{
\sigma_2^{(s)2}
}
\Bigg)
\end{gathered}
\]
for each $j$ 
\begin{math}
    \in \{2,\ldots, \tau_1^{(s)}-1\}
\end{math}. 

    \hspace{\algorithmicindent}\textbf{(iii)}  Compute 
\begin{equation}
p^{(s)}_{2}\!\left(j\right) =
\frac{
\displaystyle
\frac{1}{
\sigma_1^{(s)\,\sum_{t=1}^ {j} m_t}
\;\sigma_2^{(s)\,\sum_{t=j+1}^T m_t}
}
\exp\Bigg[
-\frac{1}{2}
D_3^* \Bigg]}{
\displaystyle
\sum_{\substack{j=2 \\ j = \tau_1^{(s)}}}^{T-1}
\frac{1}{
\sigma_1^{(1)\,\sum_{t=1}^ {j} m_t}
\;\sigma_2^{(1)\,\sum_{t=j+1}^T m_t}
}
\exp\Bigg[
-\frac{1}{2}
D_3^* \Bigg]
}
\end{equation}
\[
\begin{gathered}
D_3^*=\Bigg(
\frac{
\sum_{t=1}^{j} \sum_{i=1}^{m_t} z_{1,t}^{(s)}
}{
\sigma_1^{(1)2}
}
+
\frac{
\sum_{t=j+1}^{T} \sum_{i=1}^{m_t} z_{2,t}^{(s)}
}{
\sigma_2^{(1)2}
}
\Bigg)
\end{gathered}
\]
for  $\tau_2$ = $\tau_1^{(s)}$

\hspace{\algorithmicindent}\textbf{(d) Update $ \tau_2$} Update the change-point by setting \[
\tau_2^{(s+1)}
=
\operatorname*{arg\,max}_{j \in \{2,\ldots,T-1\}}
p^{(s)}_{2}\!\left(j\right).
\]
\newcommand{\HIDDENENDFOR}{%
  {%
    \renewcommand{\algorithmicendfor}{}%
    \ENDFOR%
  }%
}

 \hspace{\algorithmicindent} \textbf{(e) Computing probability distribution of  $\tau_1$}:
 
 \hspace{\algorithmicindent} \textbf{(i) Compute} 
\begin{equation}
p^{(s)}_{1}\!\left(j\right) =
\frac{
\displaystyle
\frac{1}{
\sigma_1^{(s)\,\sum_{t=1}^ {\tau_{2}^{(s+1)}} m_t}
\;\sigma_2^{(s)\,\sum_{t=\tau_2^{(s+1)}+1}^T m_t}
}
\exp\Bigg[
-\frac{1}{2}
D_1 \Bigg]}{
\displaystyle
\sum_{\substack{j=2 \\ j < \tau_2^{(s+1)}}}^{T-1}
\frac{1}{
\sigma_1^{(s)\,\sum_{t=1}^ {\tau_{2}^{(s+1)}} m_t}
\;\sigma_2^{(s)\,\sum_{t=\tau_2^{(s+1)}+1}^T m_t}
}
\exp\Bigg[
-\frac{1}{2}
D_1 \Bigg]
}
\end{equation}
\[
\begin{gathered}
D_1=\Bigg(
\frac{
\sum_{t=1}^{j} \sum_{i=1}^{m_t}
z_{1,t}^{(s)}
+
\sum_{t=j+1}^{\tau_{2}^{(s+1)}} \sum_{i=1}^{m_t}
z_{2,t}^{(s)}
}{
\sigma_1^{(s)2}
}
+ 
\frac{
\sum_{t=\tau_2^{(s+1)}+1}^{T} \sum_{i=1}^{m_t}
z_{1,t}^{(s)}
}{
\sigma_2^{(s)2}
}
\Bigg)
\end{gathered}
\]
for each $j$ 
\begin{math}
    \in \{2,\ldots, \tau_2^{(s+1)}-1\}
\end{math}

    \hspace{\algorithmicindent} \textbf{(ii) Compute}
\begin{equation}
p^{(s)}_{1}\!\left(j\right) =
\frac{
\displaystyle
\frac{1}{
\sigma_1^{(s)\,\sum_{t=1}^ {\tau_{2}^{(s+1)}} m_t}
\;\sigma_2^{(s)\,\sum_{t=\tau_2^{(s+1)}+1}^T m_t}
}
\exp\Bigg[
-\frac{1}{2}
D_2 \Bigg]}{
\displaystyle
\sum_{\substack{j=2 \\ j > \tau_2^{(s+1)}}}^{T-1}
\frac{1}{
\sigma_1^{(s)\,\sum_{t=1}^ {\tau_{2}^{(s+1)}} m_t}
\;\sigma_2^{(s)\,\sum_{t=\tau_2^{(s+1)}+1}^T m_t}
}
\exp\Bigg[
-\frac{1}{2}
D_2 \Bigg]
}
\end{equation}
\[
\begin{gathered}
D_2=\Bigg(
\frac{
\sum_{t=1}^{\tau_2^{(s+1)}} \sum_{i=1}^{m_t}
z_{1,t}^{(s)}}{
\sigma_1^{(s)2}
}
+ 
\frac{
\sum_{t=\tau_2^{(s+1)}+1}^{j} \sum_{i=1}^{m_t}
z_{1,t}^{(s)} +   \sum_{t=j+1}^{T} \sum_{i=1}^{m_t}
z_{2,t}^{(s)}
}{
\sigma_2^{(s)2}
}
\Bigg)
\end{gathered}
\]
for each $j$ 
\begin{math}
    \in \{\tau_{2}^{(s+1)},\ldots, T-1\}
\end{math}. 

    \hspace{\algorithmicindent} \textbf{(iii)
     Compute }
\begin{equation}
p^{(s)}_{1}\!\left(j\right) =
\frac{
\displaystyle
\frac{1}{
\sigma_1^{(s)\,\sum_{t=1}^ {\tau_{2}^{(s+1)}} m_t}
\;\sigma_2^{(s)\,\sum_{t=\tau_2^{(s+1)}+1}^T m_t}
}
\exp\Bigg[
-\frac{1}{2}
D_3 \Bigg]}{
\displaystyle
\sum_{\substack{j=2 \\ j = \tau_2^{(s+1)}}}^{T-1}
\frac{1}{
\sigma_1^{(s)\,\sum_{t=1}^ {\tau_{2}^{(s+1)}} m_t}
\;\sigma_2^{(s)\,\sum_{t=\tau_2^{(s+1)}+1}^T m_t}
}
\exp\Bigg[
-\frac{1}{2}
D_3 \Bigg]
}
\end{equation}
\[
\begin{gathered}
D_3=\Bigg(
\frac{
\sum_{t=1}^{j} \sum_{i=1}^{m_t}
z_{1,t}^{(s)}}{
\sigma_1^{(s)2}
}
+
\frac{
\sum_{t=j+1}^{T} \sum_{i=1}^{m_t}
z_{2,t}^{(s)}  
}{
\sigma_2^{(s)2}
}
\Bigg)
\end{gathered}
\]

for  $j$ = $\tau_2^{(s+1)}$

\hspace{\algorithmicindent}\textbf{(f) Update $ \tau_1$} Update the change-point by setting \[
\tau_1^{(s+1)}
=
\operatorname*{arg\,max}_{j \in \{2,\ldots,T-1\}}
p^{(s)}_{1}\!\left(j\right).
\]
    \hspace{\algorithmicindent}\textbf{(g) Parameter Update:} Given the values of  $\tau_2^{(s+1)}$ and $\tau_1^{(s+1)}$, update $\mu_1^{(s+1)}, \mu_2^{(s+1)}, \Psi^{(s+1)},  \sigma_1^{(s+1)}, \sigma_2^{(s+1)}, K_\epsilon^{(s+1)},\{\alpha_t^{(s+1)}\}_{t=1}^{T}$ by sampling from their full conditional distributions.

\STATE \textbf{Output:} Posterior samples of the parameters under study.
\end{algorithmic}
\kern2pt\hrule

\begin{algorithm}[H]
\caption{Iterative Gibbs Sampling for Change-point Detection in the Mean Function of a Functional Time Series }
\label{alg:changepoint1}
\begin{algorithmic}[1]

\STATE \textbf{Initialization:} Set initial values $\tau^{(1)}$ and hence
$\mu_1^{(1)}$, $\mu_2^{(1)}$, $\{\alpha_t^{(1)}\}_{t=1}^{T}$,
$\sigma_\nu^{(1)}$, $\Psi^{(1)}$, and $K_\epsilon^{(1)}$ as in Section ~\eqref{initialization}.
Then, re-estimate $\{\alpha_t^{(1)}\}_{t=1}^{T}$ using the robust Kalman filter.

\STATE \textbf{For} $s=1,\ldots,S$ \textbf{do}

 \hspace{\algorithmicindent}\textbf{(a) Computing probability distribution of $\tau$:}
For each candidate $j\in\{2,\ldots,T-1\}$, compute
\begin{equation}
\label{eqn9}
p^{(s)}\!\left(j\right)
=
\frac{\exp\left\{-\frac{1}{2\sigma_\nu^{(s)2}}\mathcal{SS}(j)\right\}}
{\sum_{k=2}^{T-1}
\exp\left\{-\frac{1}{2\sigma_\nu^{(s)2}}\mathcal{SS}(k)\right\}}.
\end{equation}

where
\[
\begin{aligned}
\mathcal{SS}(j)
&=
\sum_{t=1}^{j}\sum_{i=1}^{m_t}
z_{1,t}^{(s)} +
\sum_{t=j+1}^{T}\sum_{i=1}^{m_t}
z_{2,t}^{(s)}
\end{aligned}
\]

\hspace{\algorithmicindent}\textbf{(b) Coordinate Update:}
Update the change-point by setting
\[
\tau^{(s+1)}
=
\operatorname*{arg\,max}_{j\in\{2,\ldots,T-1\}}
p^{(s)}\!\left(j\right).
\]

\hspace{\algorithmicindent}\textbf{(c) Parameter Update:}
Given $\tau^{(s+1)}$, update the remaining parameters
$\boldsymbol{\mu}_1^{(s+1)}$,
$\boldsymbol{\mu}_2^{(s+1)}$,
$\mathbf{\Psi}^{(s+1)}$,
$\sigma_\nu^{(s+1)}$,
$\mathbf{K}_\epsilon^{(s+1)}$, and
$\{\boldsymbol{\alpha}_t^{(s+1)}\}_{t=1}^{T}$
by sampling from their respective full conditional distributions.

\STATE \textbf{Output:} Posterior samples.

\end{algorithmic}
\end{algorithm}

\begin{algorithm}[H]
\caption{Iterative Gibbs Sampling for Change-point Detection in the Observational noise variance of a Functional Time Series }
\label{alg:changepoint2}
\begin{algorithmic}[1]

\STATE \textbf{Initialization:} Set the initial value $\tau^{(1)}$ and hence estimate
$\mu^{(1)}$, $\{\alpha_t^{(1)}\}_{t=1}^{T}$,
$\sigma_1^{(1)}$, $\sigma_2^{(1)}$,
$\Psi^{(1)}$, and $K_\epsilon^{(1)}$ as in Section ~\eqref{initialization}.
Then, re-estimate $\{\alpha_t^{(1)}\}_{t=1}^{T}$ using the Kalman smoother.

\STATE \textbf{For} $s=1,\ldots,S$ \textbf{do}

 \hspace{\algorithmicindent}\textbf{(a) Computing probability distribution of $\tau$:}
For each candidate $j\in\{2,\ldots,T-1\}$, compute
\begin{equation}
\label{10}
p^{(s)}\!\left(j\right)
=
\frac{q(j)}
{\sum_{k=2}^{T-1} q(k)},
\end{equation}

Define
\[
 z_{t}^{(s)}=\left(y_{i,t}-\mu^{(s)}(k_{i,t})-\alpha_t^{(s)}(k_{i,t})\right)^2 
\]
Then,
\[
q(j)
=
\frac{1}
{\sigma_1^{(s)\sum_{t=1}^{j}m_t}}
\frac{1}
{\sigma_2^{(s)\sum_{t=j+1}^{T}m_t}}
\exp\left\{
-\frac{1}{2}
\left[
\frac{\displaystyle\sum_{t=1}^{j}\sum_{i=1}^{m_t}
z_{t}^{(s)}}
{\sigma_1^{(s)2}}
+
\frac{\displaystyle\sum_{t=j+1}^{T}\sum_{i=1}^{m_t}
z_{t}^{(s)}}
{\sigma_2^{(s)2}}
\right]
\right\}.
\]

 \hspace{\algorithmicindent}\textbf{(b) Coordinate Update:}
Update the change-point by setting
\[
\tau^{(s+1)}
=
\operatorname*{arg\,max}_{j\in\{2,\ldots,T-1\}}
p^{(s)}\!\left(j\right).
\]

 \hspace{\algorithmicindent}\textbf{(c) Parameter Update:}
Given $\tau^{(s+1)}$, update the remaining parameters
$\mu^{(s+1)}$,
$\Psi^{(s+1)}$,
$\sigma_1^{(s+1)}$,
$\sigma_2^{(s+1)}$,
$K_\epsilon^{(s+1)}$,
and
$\{\alpha_t^{(s+1)}\}_{t=1}^{T}$
by sampling from their respective full conditional distributions.

\STATE \textbf{Output:} Posterior samples of the parameters under study.

\end{algorithmic}
\end{algorithm}

\vspace{12pt}

Algorithm for detecting change point only in the AR operator follows Step ~2(a),(b) of ~\eqref{algo7}. Algorithm for simultaneous shift in mean and volatility follow Step ~2(c)-(g) of ~\eqref{algo7} whereas for shift in mean and operator Step ~2(a),(b) of ~\eqref{algo7} along with ~\eqref{alg:changepoint1} is used. For shift in volatility and operator, Step ~2(a),(b) of ~\eqref{algo7} along with ~\eqref{alg:changepoint2} is used.

A critical challenge in regime-switching processes is the structural hierarchy between the latent state evolution and the realized observations. When a change point occurs at the operator or evolution level, it alters the underlying latent mechanism that governs the system's dynamics. In contrast, shifts in the mean or volatility occur directly at the observational level. Because we only have access to the observed data sequence rather than the latent states themselves, detecting a change in the operator requires the system to first transition through this latent layer and manifest its effects in the data. Consequently, a change originating at the deeper evolution level inherently demands a larger sample size and a longer observation window to transcend into the visible data space, compared to sudden, explicit shifts that occur directly at the observational level.

\section{SIMULATION STUDIES}{\label{sim}}
%In this section, we shall 
In this section, we shall provide results from different simulation studies under various modeling setups.

\subsection{\MakeUppercase{Detection of Simultaneous Distortion}} \label{sim1}
In this part, we shall talk about the most relevant model in time; mean-volatility-AR operator change point model. 

Simulation results are provided for T=500, m=30 under all possible combinations of ($\mu_1(u),\mu_2(u)$) = ($\frac{1}{10} u^3 \sin(2\pi u)$, $\frac{1}{10} u^3 \sin(2\pi u)+u^{2}$) and ($sin(u),cos(u)$) as well as $(\sigma_1,\sigma_2)$ = $(0.002,0.02)$ and $\psi_1(u, v) \propto \frac{0.75}{\pi(0.3)(0.4)} \exp\left\{ -\frac{(u - 0.2)^2}{(0.3)^2} - \frac{(v - 0.3)^2}{(0.4)^2}\right\} + \frac{0.45}{\pi(0.3)(0.4)} \exp\left\{-\frac{(u - 0.7)^2}{(0.3)^2} - \frac{(v - 0.8)^2}{(0.4)^2}\right\}$ and $\psi_2(u, v) \propto v$ rescaled according to a pre-specified squared norm, $C_{\psi_i} = \int \int \psi^2_i(u, v) \,du \,dv $, with $C_{\psi_1} = 0.8$ and $C_{\psi_2} = 0.01$ respectively. We use the covariance function parameterization $K_{\epsilon} = \sigma^2 R_{\rho}$, where $R_{\rho}$ is the \cite{matern2013spatial} correlation function:

\begin{equation*}
R_{\rho}(u, v) = \frac{1}{2^{\rho_1-1} \Gamma(\rho_1)} \left( \frac{\|u - v\|}{\rho_2} \right)^{\rho_1} K_{\rho_1} \left( \frac{\|u - v\|}{\rho_2} \right),
\end{equation*}

where $\Gamma(\cdot)$ is the gamma function, $K_{\rho_1}$ is the modified Bessel function of order $\rho_1$, and $\rho = (\rho_1, \rho_2)$ are the parameters (Matérn, 2013). We let $\sigma = 0.01$ and $\rho = (\rho_1, 0.1)$, with $\rho_1 = 2.5$ for smooth (twice-differentiable) sample paths. Table ~\eqref{tab:merged_break_results} depicts the simultaneous changepoints under one such scenario.

\begin{table}[H]
\centering
\renewcommand{\arraystretch}{2.5} % Balanced space for fractions without over-stretching
\caption{Estimated change points $(\hat{\tau}_1, \hat{\tau}_2, \hat{\tau}_3)$ under varying true break configurations.}
\label{tab:merged_break_results}
\begin{tabular}{|l | c c c|}
\hline
\diagbox[width=7em]{\raisebox{-0.5ex}{$\tau_1$}}{\raisebox{0.5ex}{$\tau_2$}} & $\lceil T/4\rceil=125$ & $\frac{T}{2}=250$ & $\lceil 3T/4\rceil=375$ \\ \hline
\hline
\multicolumn{4}{|c|}{\textbf {$\tau_3 = \lceil T/4\rceil=125$}} \\ \hline
$\lceil T/4\rceil$ & (125,126,125) & (125,250,125) & (125,375,125) \\ 
$\frac{T}{2}$  & (250,126,118) & (250,250,125) & (250,375,146) \\ 
$\lceil 3T/4\rceil$ & (375,125,110) & (375,250,119) & (375,375,126) \\ \hline
\hline
\multicolumn{4}{|c|}{\textbf {$\tau_3 = \frac{T}{2}=250$}} \\ \hline
$\lceil T/4\rceil$  & (125,125,250) & (125,250,250) & (125,375,250) \\  
$\frac{T}{2}$  & (250,127,250) & (250,250,250) & (250,375,235) \\  
$\lceil 3T/4\rceil$ & (375,125,300) & (375,250,221) & (375,375,298) \\ \hline
\hline
\multicolumn{4}{|c|}{\textbf {$\tau_3 = \lceil 3T/4\rceil=375$}} \\ \hline
$\lceil T/4\rceil$ & (125,125,379) & (125,250,374) & (125,375,379) \\  
$\frac{T}{2}$  & (250,125,337) & (250,250,375) & (250,375,375) \\  
$\lceil 3T/4\rceil$ & (375,125,389) & (375,250,375) & (375,375,375) \\ \hline
\end{tabular}
\end{table}
It has been observed that across all quartiles, the estimated change points align with the true ones.

\subsection{\MakeUppercase{Detection of Distortion in the AR operator}}
In this part, we shall demonstrate the detection of change in the latent variable process.
For all simulations, the mean function is $\mu(u) = \frac{1}{10} u^3 \sin(2\pi u)$ and the measurement errors are identically distributed  $\nu_{i,t} \overset{\text{iid}}{\sim} N(0, \sigma^2_\nu)$ with $\sigma_\nu = 0.002$ . As mentioned in Section \ref{algo}, we need a larger sample in this case than in the previous cases. Accordingly we vary the sample size from small $T = 350$ to large $T = 500$ . The FAR(1) kernels used  are the Bimodal-Gaussian kernel, $\psi_1(u, v) \propto \frac{0.75}{\pi(0.3)(0.4)} \exp\left\{ -\frac{(u - 0.2)^2}{(0.3)^2} - \frac{(v - 0.3)^2}{(0.4)^2}\right\} + \frac{0.45}{\pi(0.3)(0.4)} \exp\left\{-\frac{(u - 0.7)^2}{(0.3)^2} - \frac{(v - 0.8)^2}{(0.4)^2}\right\}$ and $\psi_2(u, v) \propto v$ rescaled according to a pre-specified squared norm, $C_{\psi_i} = \int \int \psi^2_i(u, v) \,du \,dv $, with $C_{\psi_i} < 1$ for stationarity. We select $C_{\psi_1} = 0.8$ for the  simulation. $K_\epsilon$ is same as in Section \eqref{sim1}. \\
Two issues have been considered here. Firstly, we want to see how the estimation result varies with the change in the Frobenius norm $||\Psi_1-\Psi_2||_{F}$. Secondly, we investigate the impact of the signal-to-noise ratio, which is  quantified in terms of $\sigma_\nu$, on the estimation result. Table \ref{tab:4} shows that with increase in the  Frobenius norm, results become more accurate. The choices of $C_{\psi_2} = 0.1$ and 0.01 corresponds to the two different Frobenius norms. Tables \ref{tab:5} clearly depicts that for accurate estimation $\sigma_\nu$ need not be too small. In many practical situations $\sigma_\nu$ may be moderately large and is not in control of the experimenter. The simulation result shows that our method performs reasonably well under such situations also.
\\
\begin{table}[H]
\renewcommand{\arraystretch}{1.9}
\centering
\large
\caption{Change in the AR-operator: $m=30$, $C_{\Psi_{1}}=0.8$, $\sigma_\nu=0.002$}
\label{tab:4}
\begin{tabular}{ccccccccc}
\toprule
\multirow{2}{*}{$T$} &
\multirow{2}{*}{$C_{\Psi_{2}}$} &
\multirow{2}{*}{$\|\Psi_1-\Psi_2\|_F$} &
\multicolumn{2}{c}{CP at $\lceil T/4\rceil$} &
\multicolumn{2}{c}{CP at $T/2$} &
\multicolumn{2}{c}{CP at $\lceil 3T/4\rceil$} \\
\cmidrule(lr){4-5}\cmidrule(lr){6-7}\cmidrule(lr){8-9}
 & & &
Actual & Estimated &
Actual & Estimated &
Actual & Estimated \\
\midrule
350 & 0.10 & 20.5836 & 88 & 94 & 175 & 165 & 263 & 264 \\
    & 0.01 & 24.2300 & 88 & 94 & 175 & 168 & 263 & 264 \\
\cmidrule(lr){2-9}
500 & 0.10 & 20.5836 & 125 & 124 &250& 249&375&377\\
& 0.01& 24.23& 125&124&250&252&375&377\\
\bottomrule
\end{tabular}
\end{table}
\begin{table}[H]
\renewcommand{\arraystretch}{1.9}
\centering
\large
\caption{Change in the AR-operator: $m=30$, $C_{\Psi_{1}}=0.8$, $\sigma_\nu=0.02$}
\label{tab:5}
\begin{tabular}{ccccccccc}
\toprule
\multirow{2}{*}{$T$} &
\multirow{2}{*}{$C_{\Psi_{2}}$} &
\multirow{2}{*}{$\|\Psi_1-\Psi_2\|_F$} &
\multicolumn{2}{c}{CP at $\lceil T/4\rceil$} &
\multicolumn{2}{c}{CP at $T/2$} &
\multicolumn{2}{c}{CP at $\lceil 3T/4\rceil$} \\
\cmidrule(lr){4-5}\cmidrule(lr){6-7}\cmidrule(lr){8-9}
 & & &
Actual & Estimated &
Actual & Estimated &
Actual & Estimated \\
\midrule
350 & 0.10 & 20.5836 & 88 & 70 & 175 & 155 & 263 & 248 \\
    & 0.01 & 24.2300 & 88 & 71 & 175 & 167 & 263 & 256 \\
\cmidrule(lr){2-9}
500 & 0.10 & 20.5836 & 125 & 113 &250& 242&375&375\\
& 0.01& 24.23& 125&113&250&242&375&375\\
\bottomrule
\end{tabular}
\end{table}

A critical, yet under-explored, scenario involves the system shifting from a state of independence to latent dependence. This essentially means that the AR-operator shifts from being a null operator to a non-null one.

Table \ref{depind} shows that our proposed algorithm can detect phase transitions almost correctly when the Frobenius norm of the AR-operator is large enough. For a smaller magnitude, changes can be detected near the centre and not away from the centre.

\begin{table}[H]
\centering
\renewcommand{\arraystretch}{1.9}
\large
\caption{Transition from Independence to Latent Dependence, $m=30$,$\Psi_1=0$, $\sigma_\nu=0.002$}
\label{depind}
\begin{tabular}{ccc|cc|cc|cc}
\toprule
\multirow{2}{*}{$T$} &
\multirow{2}{*}{$C_{\Psi_{2}}$} &
\multirow{2}{*}{$\|\Psi_1-\Psi_2\|_F$} &
\multicolumn{2}{c|}{CP at $\left\lceil\frac{T}{4}\right\rceil$} &
\multicolumn{2}{c|}{CP at $\frac{T}{2}$} &
\multicolumn{2}{c}{CP at $\left\lceil\frac{3T}{4}\right\rceil$} \\
\cmidrule(lr){4-5}
\cmidrule(lr){6-7}
\cmidrule(lr){8-9}
&&&
Actual & Estimated &
Actual & Estimated &
Actual & Estimated \\
\midrule
\multirow{2}{*}{350}
& 0.1*
& 9.56
& 88 & 274
& 175 & 188
& 263 & 155 \\
\cmidrule(lr){2-9}
& 0.8**
& 26.30
& 88 & 92
& 175 & 175
& 263 & 266 \\
\midrule
\multirow{2}{*}{500}
& 0.1*
& 9.56
& 125 & 54
& 250 & 255
& 375 & 413 \\
\cmidrule(lr){2-9}
& 0.8**
& 26.30
& 125 & 125
& 250 & 250
& 375 & 379 \\
\bottomrule
\end{tabular}
\end{table}

\begingroup
\renewcommand{\thefootnote}{*}
\footnotetext{Indicates the bivariate Gaussian kernel
\[
\psi_1(u,v)\propto
\frac{0.75}{\pi(0.3)(0.4)}
\exp\!\left\{
-\frac{(u-0.2)^2}{(0.3)^2}
-\frac{(v-0.3)^2}{(0.4)^2}
\right\}
+
\frac{0.45}{\pi(0.3)(0.4)}
\exp\!\left\{
-\frac{(u-0.7)^2}{(0.3)^2}
-\frac{(v-0.8)^2}{(0.4)^2}
\right\}.
\]}
\endgroup

\begingroup
\renewcommand{\thefootnote}{**}
\footnotetext{Indicates the kernel $\psi_2(u,v)\propto v.$}
\endgroup

\subsection{\MakeUppercase{Detection of Mean Shifts and Volatility Change}}

Consider the sample size varying from small $T = 50$ to large $T = 200$ and the grid size $m=30$. All possible combinations of ($\mu_1(u),\mu_2(u)$) = ($\frac{1}{10} u^3 \sin(2\pi u)$, $\frac{1}{10} u^3 \sin(2\pi u)+u^{2}$) and ($sin(u),cos(u)$) as well as $(\sigma_1,\sigma_2)$ = $(0.002,0.01)$ and $(0.002,0.02)$ have been considered. The operator $\Psi$ is the $\Psi_1$ operator and $K_\epsilon$ is the same as in Section \eqref{sim1}. An illustration for varying sample size has been provided in Table \ref{tab:3}.

\vspace{2mm}
\begingroup
\renewcommand\thefootnote{}\footnote{$\lceil p \rceil$ denotes the smallest integer greater than or equal to $p$.}
\addtocounter{footnote}{-1}
\endgroup

\begin{table}[H]
\renewcommand{\arraystretch}{2.5}
    \centering
    
    \caption{Change in both mean and volatility: $\mu_1(u) = \frac{1}{10} u^3 \sin(2\pi u)$, $\mu_2(u)=\mu_1(u)+u^2$,$\sigma_1=0.002,\sigma_2=0.02$}
    \label{tab:3}
    
    % Adjust horizontal padding between columns to ensure it fits the page
    \setlength{\tabcolsep}{10pt} 
    \small % Slightly smaller font size for 10-column layout
    
    \begin{tabular}{l ccc ccc ccc}
        \toprule
        & \multicolumn{3}{c}{\textbf{T = 50}} & \multicolumn{3}{c}{\textbf{T = 100}} & \multicolumn{3}{c}{\textbf{T = 200}} \\
        \cmidrule(lr){2-4} \cmidrule(lr){5-7} \cmidrule(lr){8-10}
        
        \diagbox[width=6em]{$\tau_1$}{$\tau_2$} 
        & $\lceil \frac{T}{4} \rceil=13$ & $\frac{T}{2}=25$ & $\lceil \frac{3T}{4} \rceil=38$ 
        & $\lceil \frac{T}{4} \rceil=25$ & $\frac{T}{2}=50$ & $\lceil \frac{3T}{4} \rceil=75$ 
        & $\lceil \frac{T}{4} \rceil=50$ & $\frac{T}{2}=100$ & $\lceil \frac{3T}{4} \rceil=150$ \\
        \midrule
        
        $\lceil \frac{T}{4} \rceil$ & (13.,13) & (13,25) & (13,38) & (25,25) & (25,50) & (25,75) & (50,50) & (50,100) & (50,150) \\
        \addlinespace[0.5em] % Adds a small gap between rows for readability
        
        $\frac{T}{2}$               & (25,13) & (25,25) & (25,38) & (50,25) & (50,50) & (50,75) & (100,50) & (100,100) & (100,150) \\
        \addlinespace[0.5em]
        
        $\lceil \frac{3T}{4} \rceil$ & (38,13) & (38,25) & (38,38) & (75,25) & (75,50) & (75,75) & (150,50) & (150,100) & (150,150)\\
        \bottomrule
    \end{tabular}
\end{table}

From the above noted results, we can infer that irrespective of the sample sizes, exact change point locations can be estimated. We have done simulations under the independent setup, ie. $\Psi$= 0. In each such cases, the mean change and the volatility shift can be detected accurately as in the dependent case.

When analyzing the same dataset, the E-Divisive method of \citet{matteson2014nonparametric} yields less reliable estimates because it fails to account for temporal dependence. Furthermore, under small sample sizes, structural distortions remain difficult to detect. As shown in Table~\ref{tab1:3}, true change point locations are frequently obscured by spurious signals.

\begin{table}[H]
\renewcommand{\arraystretch}{2.5}
    \centering
    
    \caption{Detection of change point with changes in both mean and volatility: $\mu_1(u) = \frac{1}{10} u^3 \sin(2\pi u)$, $\mu_2(u)=\mu_1(u)+u^2$,$\sigma_1=0.002,\sigma_2=0.02$ using E-Divisive of \cite{matteson2014nonparametric}}
    \label{tab1:3}
    
    % Adjust horizontal padding between columns to ensure it fits the page
    \setlength{\tabcolsep}{5pt} 
    \small % Slightly smaller font size for 10-column layout
    
    \begin{tabular}{l ccc ccc ccc}
        \toprule
        & \multicolumn{3}{c}{\textbf{T = 50}} & \multicolumn{3}{c}{\textbf{T = 100}} & \multicolumn{3}{c}{\textbf{T = 200}} \\
        \cmidrule(lr){2-4} \cmidrule(lr){5-7} \cmidrule(lr){8-10}
        
        \diagbox[width=6em]{$\tau_1$}{$\tau_2$} 
        & $\lceil \frac{T}{4} \rceil=13$ & $\frac{T}{2}=25$ & $\lceil \frac{3T}{4} \rceil=38$ 
        & $\lceil \frac{T}{4} \rceil=25$ & $\frac{T}{2}=50$ & $\lceil \frac{3T}{4} \rceil=75$ 
        & $\lceil \frac{T}{4} \rceil=50$ & $\frac{T}{2}=100$ & $\lceil \frac{3T}{4} \rceil=150$ \\
        \midrule
        
        $\lceil \frac{T}{4} \rceil$ & No & No & No & 30 & 30,62 & 30,62 & 50,163,107 & 50,125,80,153 & 50,113,155,82 \\
        \addlinespace[0.5em] % Adds a small gap between rows for readability
        
        $\frac{T}{2}$               & No & No & No & 50 & 50 & 50 & 100,63,158 & 100,43,167,130 & 100,43,155 \\
        \addlinespace[0.5em]
        
        $\lceil \frac{3T}{4} \rceil$ & No & No & No & 70,32 & 70,40 & 70,40 & 150,114,51,83 & 150,116,43,73 & 150,59,93\\
        \bottomrule
    \end{tabular}
\end{table}
%\textcolor{violet}{tab:3 has been used to label both tab2.tex and tab3.tex which is creating confusion.} \textcolor{blue}{resolved}

\subsection{\MakeUppercase{Discussions}}
\begin{figure}[H]
    \centering
    \includegraphics[width=0.80\linewidth]{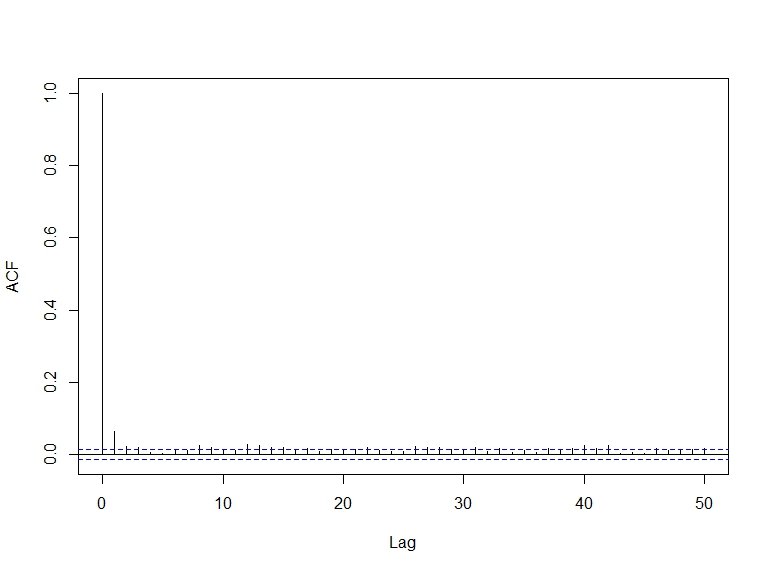}
    \caption{ACF plot of $\hat{K_\epsilon}$[5,9] across iterations for T=500, change point at 375, $\sigma_\nu$=0.002, $C_{\Psi_{1}}=0.8$, $C_{\Psi_{2}}=0.1$}
    \label{fig:keps}
\end{figure}

\begin{figure}[H]
    \centering
    \includegraphics[width=0.80\linewidth]{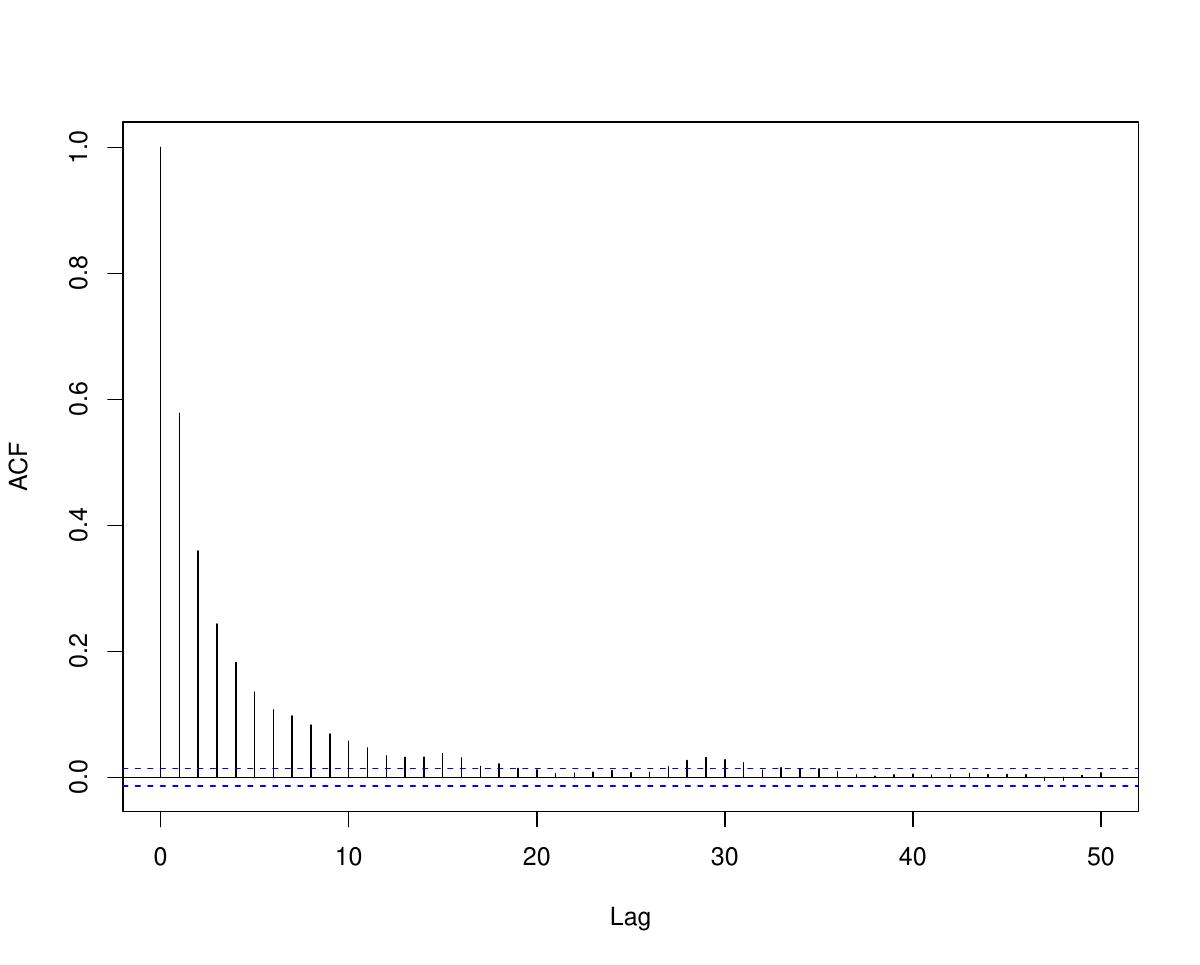}
    \caption{ACF plot of $\hat{\tau}$ across iterations for T=500, change point at 375, $\sigma_\nu$=0.002, $C_{\Psi_{1}}=0.8$, $C_{\Psi_{2}}=0.1$}
    \label{fig:tau}
\end{figure}

\begin{figure}[H]
\centering
\begin{overpic}[width=0.55\textwidth]{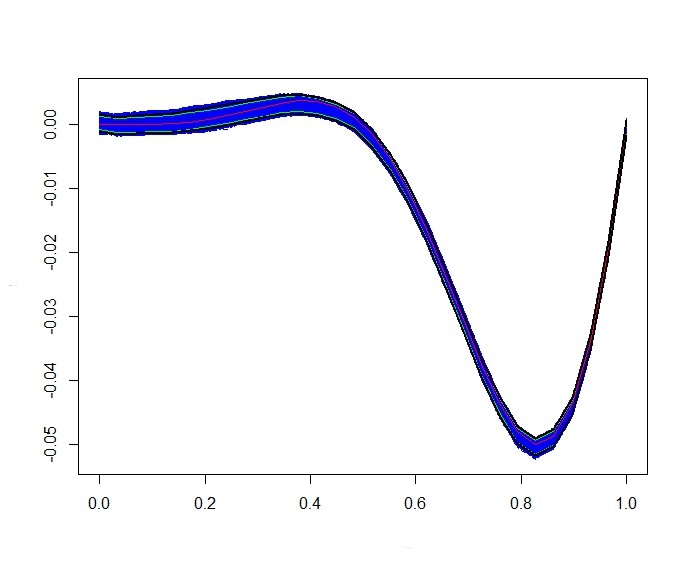}
    
    % X-Axis Label (Horizontal position: 45%, Vertical position: 0%)
    \put (45, 5) {Grid Points}
    
    % Y-Axis Label (Horizontal position: 1%, Vertical position: 50%)
    \put (1, 40) {$\hat{\mu}$}
    
  \end{overpic}
  \caption{95\% HPD credible interval containing the true $\mu$ for T=500 with change point at 375, $\sigma_\nu$=0.002, $C_{\Psi_1}=0.8$ and $C_{\Psi_2}=0.1$}. 
  \label{mu}
\end{figure}

\begin{figure}[H]
\centering
\begin{overpic}[width=0.55\textwidth]{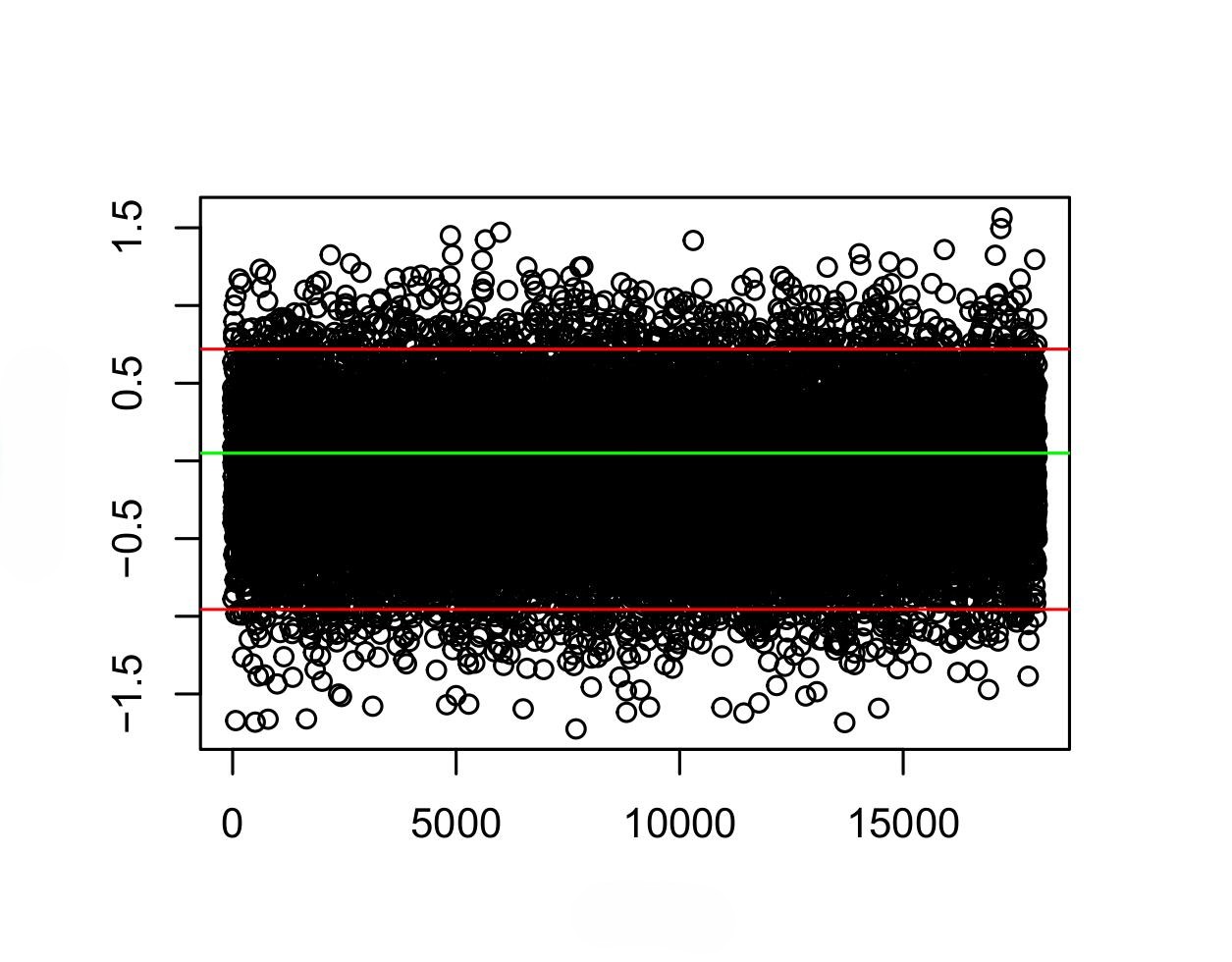}
    
    % X-Axis Label (Horizontal position: 45%, Vertical position: 0%)
    \put (45, 5) {Index}
    
    % Y-Axis Label (Horizontal position: 1%, Vertical position: 50%)
    \put (-10, 40) {$\hat{\Psi}_{1}[1,30]$}
    
  \end{overpic}
  \caption{95\% HPD credible interval containing the true ${\Psi}_{1}[1,30]$ for T=500 with change point at 375, $\sigma_\nu$=0.002, $C_{\Psi_1}=0.8$ and $C_{\Psi_2}=0.1$ }.   
  \label{psi1}
\end{figure}

\begin{figure}[H]
\centering
\begin{overpic}[width=0.55\textwidth]{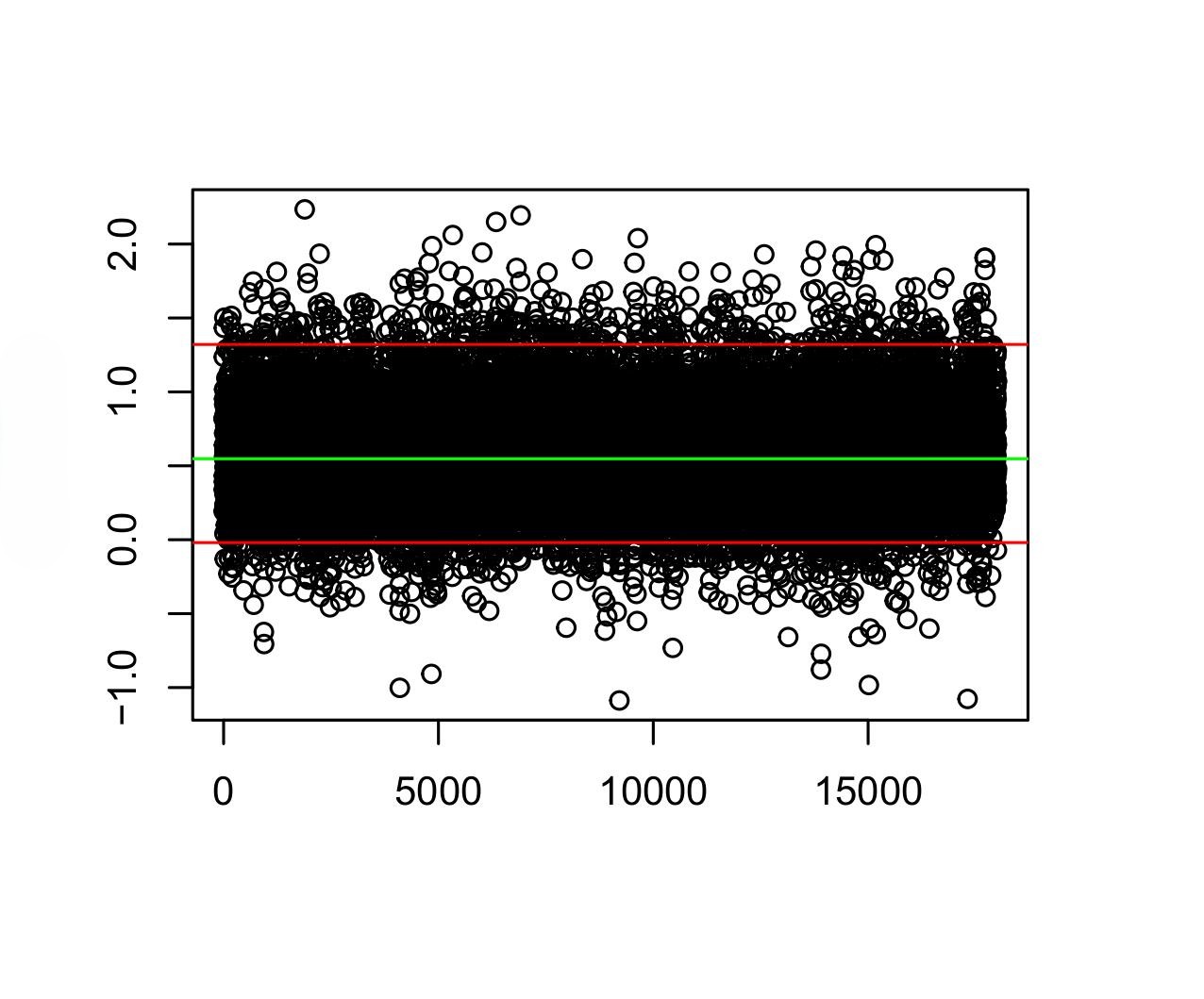}
    
    % X-Axis Label (Horizontal position: 45%, Vertical position: 0%)
    \put (45, 5) {Index}
    
    % Y-Axis Label (Horizontal position: 1%, Vertical position: 50%)
    \put (-10, 45) {$\hat{\Psi}_{2}[1,30]$}
    
  \end{overpic}
  \caption{95\% HPD credible interval containing the true ${\Psi}_{2}[1,30]$ for T=500 with change point at 375, $\sigma_\nu$=0.002, $C_{\Psi_1}=0.8$ and $C_{\Psi_2}=0.1$}.   
  \label{psi2}
\end{figure}

\begin{figure}[H]
    \centering
    \includegraphics[width=0.80\linewidth]{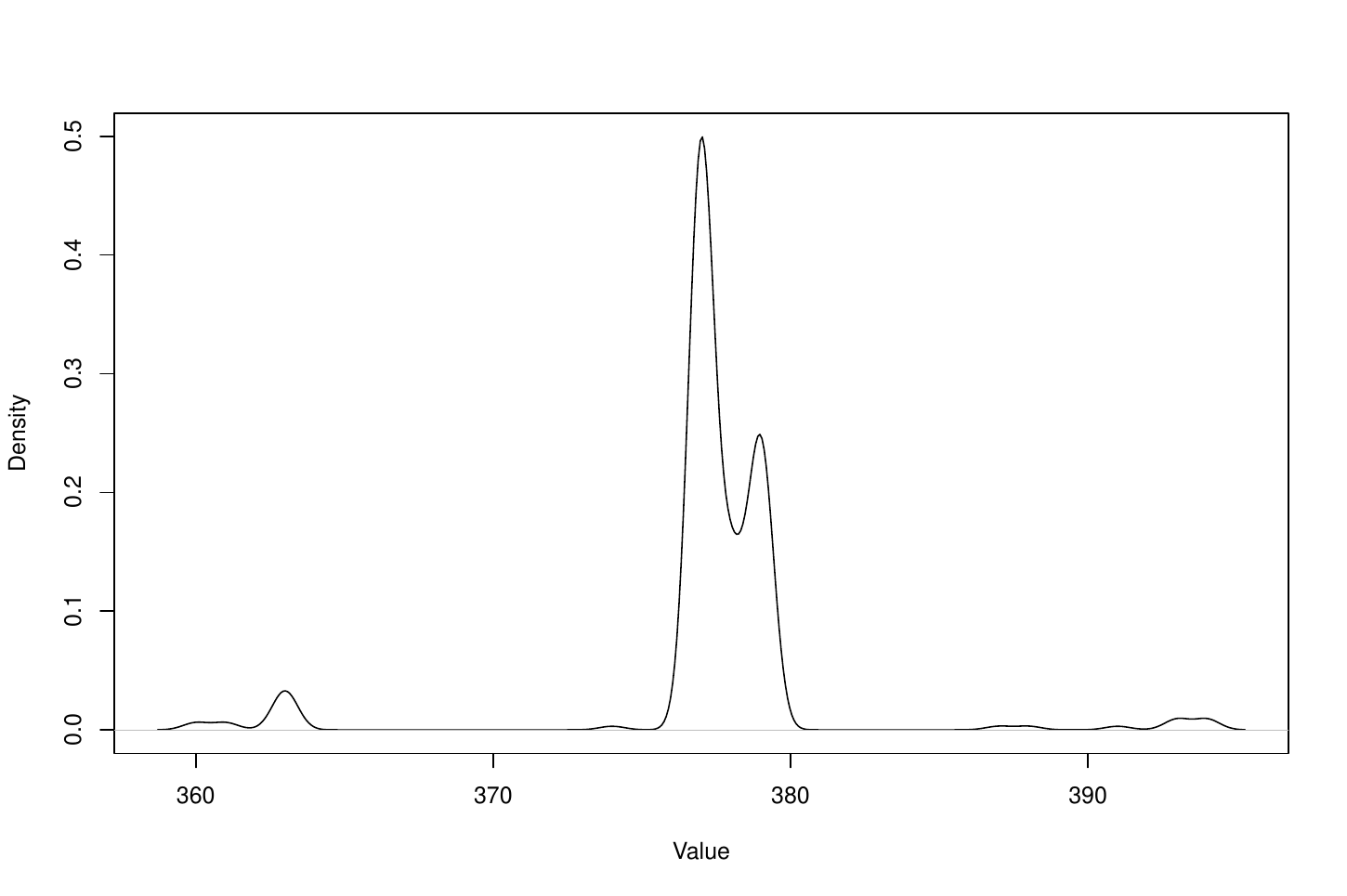}
    \caption{Kernel density plot of $\hat{\tau}$ across iterations for T=500, change point at 375, $\sigma_\nu$=0.002, $C_{\Psi_{1}}=0.8$, $C_{\Psi_{2}}=0.1$}
    \label{fig:tau1}
\end{figure}

To assess the convergence of the Gibbs sampler, we repeated the algorithm over multiple simulated datasets and monitored its performance across a large number of iterations. Overall, the findings are consistent with the reported results, with only negligible differences across repeated trials. Such variations are reflected in slight changes in the estimated probabilities across trials. \\
In a Markov chain, consecutive samples are typically correlated, as indicated by the autocorrelation function (ACF) plot. Such plots are shown in Figures ~\eqref{fig:keps} and~\eqref{fig:tau}. To reduce this serial dependence and obtain approximately independent samples, the chain is thinned by retaining only observations separated by an appropriate lag for which the autocorrelation is negligible. For, the operator problem, we have considered 20,000 iterations with a burn-in of first 2,000 observations. For all other models, 5000 iterations have been carried forward with a burn-in period of initial 2000 observations. Figure ~\eqref{mu}, ~\eqref{psi1}, ~\eqref{psi2} gives a 95 \% Highest Posterior Density Confidence Interval containing the true  $\mu$, $\Psi_1[1,30]$ and $\Psi_2[1,30]$ respectively under T=500, change point at 375, $\sigma_\nu$=0.002 and $\mu(u)$ = $\frac{1}{10} u^3 \sin(2\pi u)$. We can clearly see that the true function (values) lies well inside the confidence band. Figure ~\eqref{fig:tau1} provide the kernel density plot of the  estimated change point $\hat{\tau}$ .

\section{EMPIRICAL APPLICATIONS}{\label{dataanalysis}}
In this section, we shall illustrate the applicability of our proposed methodology for the detection of structural breaks in real data. For this purpose, three datasets are considered: the closing prices of Bitcoin, the land surface monthly average temperature, and the closing prices of the NIFTY 50 index. The analysis is mainly from two different aspects. Firstly, to show how our method works in contrast to the existing literature. Secondly, how the occurrence of significant global events has affected the market structure and whether our method do justice in acknowledging that fact.
\subsection{BITCOIN}
First, we consider the Bitcoin dataset. The period between 2020 and 2022 was arguably the most volatile and transformative era in Bitcoin's history. It moved from a niche digital asset into a mainstream institutional financial instrument, experiencing both its greatest bull run and a severe \enquote{crypto winter}. This era serves as the perfect scenario for the study of structural shifts. Some notable events in the history of Bitcoin in those years are as follows.

\textbf{The China Ban:} In May, China's strict crackdown on crypto mining and trading caused a flash crash, wiping out nearly 50\% of Bitcoin’s value in weeks. While it caused a -50\% \enquote{flash crash} in the short term, its long-term effect was actually a fundamental strengthening of the network's resilience and decentralization.

Bitcoin hit its then-all-time high of approximately \$69{,}000 in November 2021, driven by the launch of the first U.S. Bitcoin Futures ETF (BITO). There was a transition from Retail-led (Reddit/Robinhood mania) to Institutional-led (Algo-trading and ETFs). The surge of Bitcoin to its November 2021 high was a landmark moment driven largely by the launch of the ProShares Bitcoin Strategy ETF (BITO), which provided regulated, futures-based access to institutional investors and mainstream wealth managers.

\textbf{Russia Ukraine War: } This investigation analyzes the influence of the Russia--Ukraine conflict ( starting around February 24, 2022) on cryptocurrency market dynamics, focusing on Bitcoin returns and transaction volume. The empirical evidence demonstrates that the conflict significantly suppressed Bitcoin trading volume. Furthermore, this adverse effect intensified during the post-invasion phase, particularly starting one week after the outbreak of the war.

Global financial markets faced heightened uncertainty driven by unexpected US inflation surprises and subsequent monetary tightening expectations. Concurrently, cryptocurrency markets experienced severe volatility during Ethereum's Merge in September 2022, which was further exacerbated by the FTX collapse in November 2022, leading to prolonged market instability (see \cite{lu2026explains}).

\begin{figure}[H]
\centering
\includegraphics[width=0.90\textwidth]{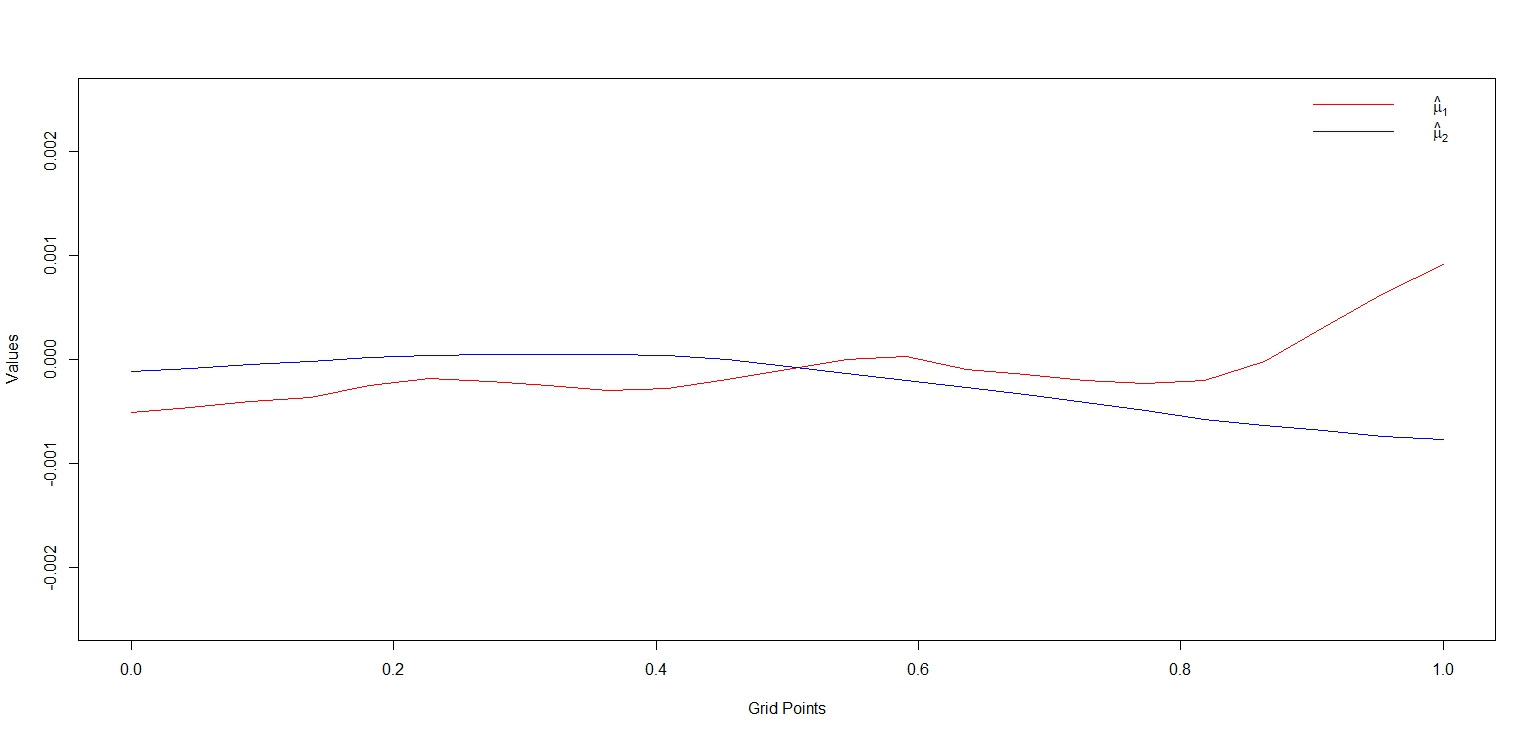}
\caption{Estimates of the pre-change mean ($\hat{\mu}_1$) and post-change mean ($\hat{\mu}_2$) for the period 2020--2022}
{\label{mean}}
\end{figure}

\begin{figure}[H]
\centering
\includegraphics[width=0.90\textwidth]{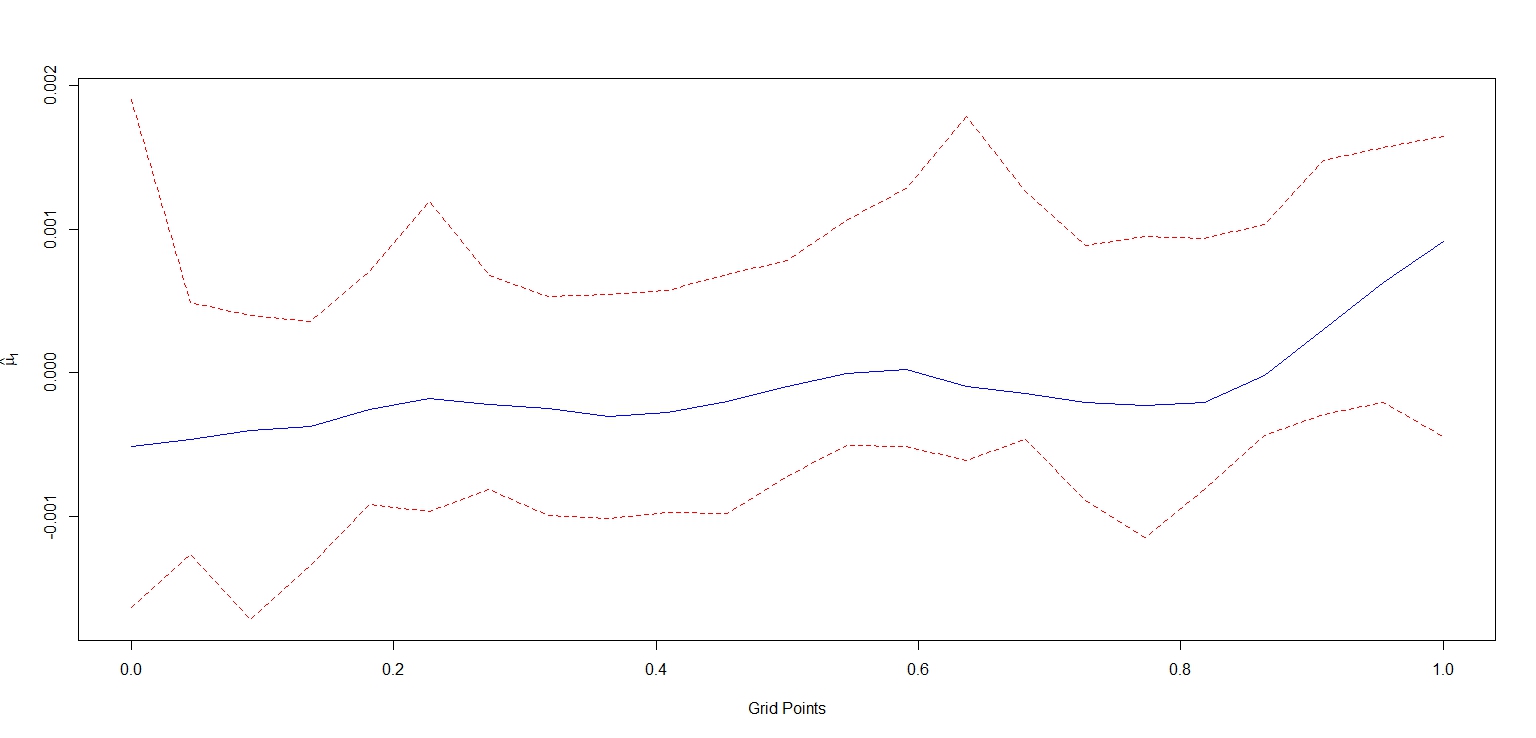}
\caption{95 \% HPD credible interval containing $\hat{\mu}_1$ for the period 2020--2022}
{\label{mu1}}
\end{figure}

\begin{figure}[H]
\centering
\includegraphics[width=0.90\textwidth]{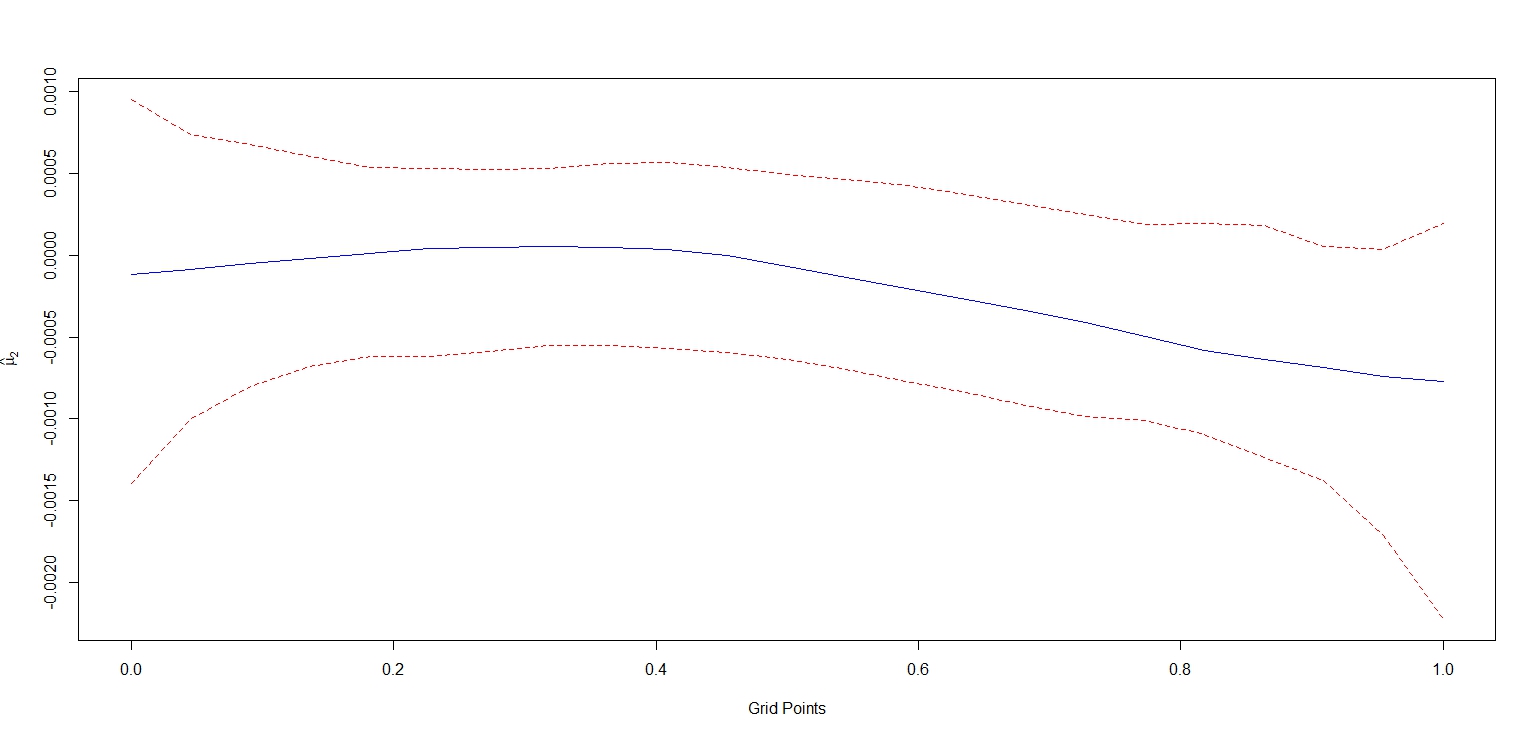}
\caption{95 \% HPD credible interval containing $\hat{\mu}_2$ for the period 2020--2022}
{\label{mu2}}
\end{figure}

\begin{figure}[H]
\centering
\includegraphics[width=0.90\textwidth]{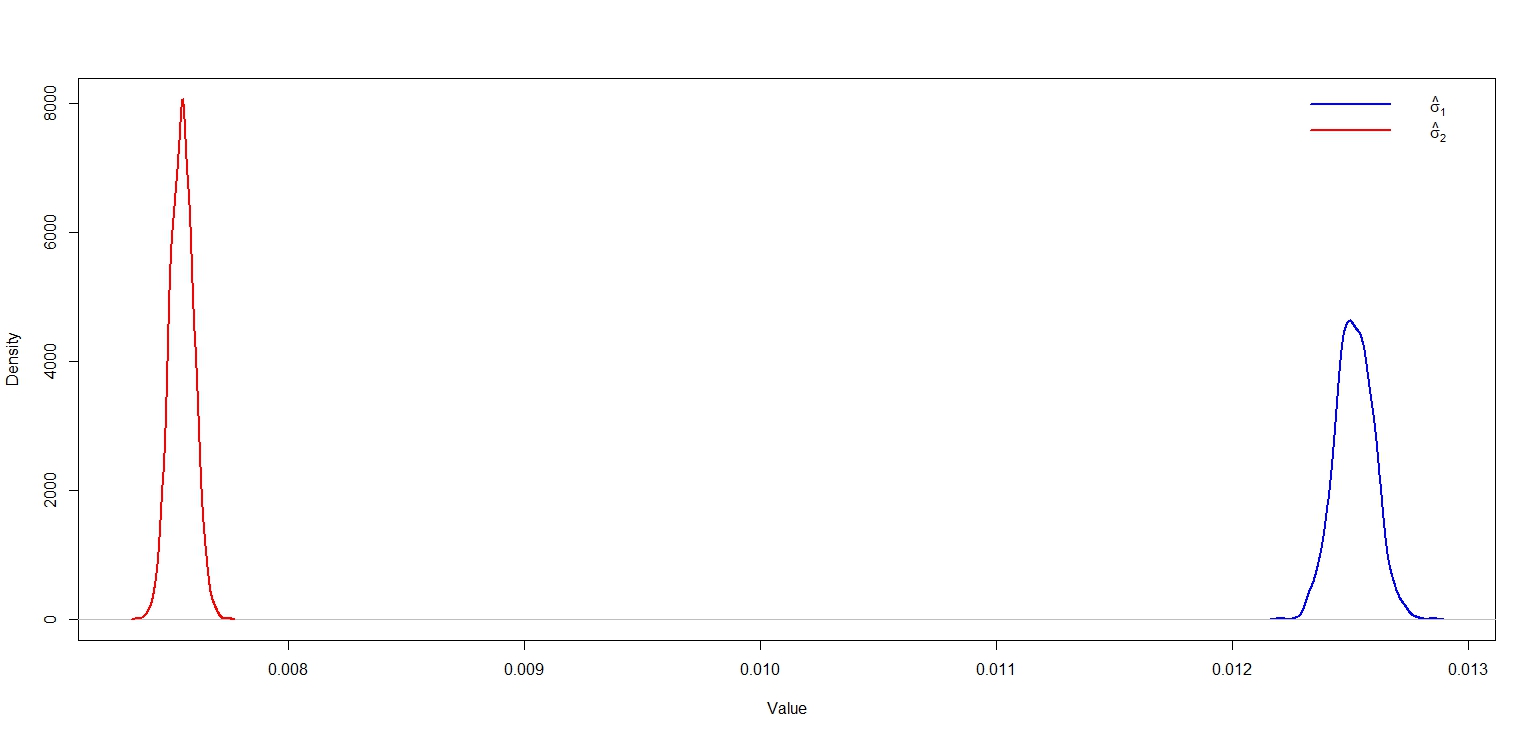}
\caption{Kernel density estimates of pre-change volatility ($\hat{\sigma}_1$) and post-change volatility ($\hat{\sigma}_2$) across iterations}
{\label{sigma}}
\end{figure}

Having obtained the insight from all this evidence, we shall proceed to our analysis. Consider the high-frequency intra-day data, precisely the hourly log returns of the closing prices for the years 2020, 2021, 2022 and 2020-2022 respectively. This can be looked upon as random curves on a grid of 23 points. Table \eqref{tab:research_data} shows the estimated change points across different models. The change in the AR-operator can only be detected on $10^{th}$ May 2022 and the pre-change as well as the post change dependence, quantified in terms of the estimated matrices $\hat{\Psi}_{1}$ and $\hat{\Psi}_{2}$ have been presented in Table ~\eqref{tab:V_values1} and ~\eqref{tab:V_values2}. The estimated norm $||\hat{\Psi}_1-\hat{\Psi}_2||_{F}$=64.58. Also, $\det (\hat{\Psi}_{1})$ $\approx$ 0 whereas $\det (\hat{\Psi}_{2})$ = 0.034. When analysing the 2020, 2021 dataset, the estimated CPs in the case of the mean, volatility simultaneous change point model exactly matches with the all parameter change point model. So, conditional on no operator change, all parameter change point model is performing like the simultaneous change point model.  Since, $19^{th}$ Dec 2020 and $11^{th}$ Jan 2021 is at the periphery of the concerned years, we shall consider them as no change positions. $23^{rd}$ March 2020 is attributable to the Covid 19 effect whereas $30^{th}$ May 2021 is linked to the China Ban. For the year 2022, estimates under all parameter change point model is more reliable than under individual or simultaneous changes. The estimated date $23^{rd}$ Sept 2022 aligns with the article of \cite{lu2026explains}. In case of 2020-2022, estimated change points are $27^{th}$ March 2022, $29^{th}$ June 2021 and $3^{rd}$ Jan 2020. Since, $3^{rd}$ Jan 2020 is at the periphery, we shall consider it as no change point. Overall, the all parameter model is most reliable among all. The change in mean aligns with the work of \cite{appiah2023impact} where the volatility change is attributable to the China Ban.  Figure ~\eqref{mean} and ~\eqref{sigma} provides the plot for the pre as well as post change mean functions as well as kernel density for pre and post change volatility for the period 2020--2022 respectively. Figure ~\eqref{mu1} and ~\eqref{mu2} provides 95\% HPD confidence interval containing $\hat{\mu}_1$ and $\hat{\mu}_2$.

\begin{table}[H]{\label{Real Table1}}
\footnotesize
\centering
\caption{Estimate of the change point locations}
\label{tab:research_data}
\renewcommand{\arraystretch}{2.5} 
\begin{tabular}{lllll}
\toprule
Model Type & 2020 & 2021 & 2022 & 2020-2022 \\
\midrule
%All parameters & ($19^{th}$ Dec, $23^{rd}$ Mar, No) & (No, $30^{th}$ May, No)& (No, $23^{rd}$ Sep, No)& ($9^{th}$ Nov'21, $1^{st}$ Jul'21, $3^{rd}$ Jan'20) \\
All parameters & ($19^{th}$ Dec, $23^{rd}$ Mar, No) & ($11^{th}$ Jan, $30^{th}$ May, No)& (No, $23^{rd}$ Sep, No)& ($27^{th}$ Mar'22, $29^{th}$ June'21, $3^{rd}$ Jan'20) \\
%AR-operator & No &$28^{th}$ May & $10^{th}$ May & $30^{th}$ May'21 \\

AR-operator & No & No & $10^{th}$ May & No \\

%Mean & $24^{th}$ Jun & $11^{th}$ May & $13^{th}$ Jul & $23^{rd}$ Sep'21 \\
%Variance & $23^{rd}$ Mar & $22^{nd}$ Jun & $23^{rd}$ Sep & $29^{th}$ Jun'21 \\

Mean,Variance & ($19^{th}$ Dec, $23^{rd}$ Mar) & ($11^{th}$ Jan, $30^{th}$ May ) & (No,$23^{rd}$ Sep ) & ($27^{th}$ Mar'22, $29^{th}$ June'21)\\
%Mean,Variance & ($19^{th}$ Dec, $23^{rd}$ Mar) & ($11^{th}$ Jan, $30^{th}$ May ) & (No,$23^{rd}$ Sep ) & ($9^{th}$ Nov 2021, $22^{nd}$ Jun'21) \\

%E-Divisive & $14^{th}$ Apr & $1^{st}$ Jun & $8^{th}$ May & $20^{th}$ Dec'20\\
\bottomrule
\end{tabular}
\end{table}

\begin{table}[H]
\centering
\caption{Pre change AR-operator $\hat{\Psi}_1$ for the year 2022}
\scriptsize
\setlength{\tabcolsep}{3pt}
\renewcommand{\arraystretch}{1.1}
\begin{tabular}{*{23}{r}}
\toprule
-1.36 & -1.32 & -1.26 & -1.34 & -1.41 & -1.37 & -1.26 & -1.17 & -1.14 & -1.12 &
-1.04 & -0.81 & -0.55 & -0.39 & -0.31 & -0.26 & -0.16 & -0.03 & 0.11 & 0.21 & 0.28 & 0.32 & 0.34 \\

-1.30 & -1.23 & -1.11 & -1.15 & -1.20 & -1.20 & -1.13 & -1.02 & -0.92 & -0.86 &
-0.81 & -0.69 & -0.54 & -0.40 & -0.30 & -0.21 & -0.08 & 0.09 & 0.27 & 0.38 & 0.40 & 0.43 & 0.54 \\

-1.12 & -1.05 & -0.98 & -1.00 & -1.01 & -0.99 & -0.95 & -0.87 & -0.74 & -0.65 &
-0.62 & -0.57 & -0.48 & -0.38 & -0.29 & -0.19 & -0.05 & 0.13 & 0.32 & 0.42 & 0.42 & 0.47 & 0.58 \\

-0.96 & -0.86 & -0.81 & -0.80 & -0.79 & -0.76 & -0.72 & -0.66 & -0.56 & -0.50 &
-0.46 & -0.40 & -0.32 & -0.28 & -0.22 & -0.14 & -0.03 & 0.10 & 0.22 & 0.33 & 0.40 & 0.45 & 0.52 \\

-0.74 & -0.67 & -0.63 & -0.60 & -0.57 & -0.55 & -0.52 & -0.46 & -0.39 & -0.35 &
-0.32 & -0.26 & -0.20 & -0.17 & -0.16 & -0.12 & -0.04 & 0.06 & 0.14 & 0.23 & 0.35 & 0.42 & 0.48 \\

-0.47 & -0.48 & -0.47 & -0.43 & -0.40 & -0.37 & -0.36 & -0.31 & -0.23 & -0.19 &
-0.19 & -0.16 & -0.11 & -0.09 & -0.09 & -0.08 & -0.02 & 0.07 & 0.12 & 0.18 & 0.27 & 0.37 & 0.48 \\

-0.21 & -0.26 & -0.32 & -0.29 & -0.25 & -0.22 & -0.21 & -0.16 & -0.06 & -0.03 &
-0.05 & -0.03 & 0.00 & 0.01 & 0.03 & 0.07 & 0.10 & 0.13 & 0.15 & 0.17 & 0.21 & 0.28 & 0.42 \\

-0.05 & -0.08 & -0.16 & -0.14 & -0.10 & -0.08 & -0.05 & 0.00 & 0.08 & 0.11 &
0.10 & 0.10 & 0.12 & 0.12 & 0.14 & 0.18 & 0.21 & 0.21 & 0.20 & 0.18 & 0.18 & 0.24 & 0.37 \\

0.01 & 0.04 & 0.03 & 0.05 & 0.07 & 0.07 & 0.13 & 0.17 & 0.19 & 0.22 &
0.25 & 0.23 & 0.20 & 0.21 & 0.19 & 0.16 & 0.20 & 0.26 & 0.25 & 0.21 & 0.20 & 0.30 & 0.40 \\

0.12 & 0.20 & 0.23 & 0.26 & 0.27 & 0.27 & 0.32 & 0.35 & 0.32 & 0.32 &
0.33 & 0.30 & 0.26 & 0.26 & 0.22 & 0.17 & 0.21 & 0.26 & 0.24 & 0.22 & 0.28 & 0.39 & 0.49 \\

0.34 & 0.40 & 0.43 & 0.46 & 0.48 & 0.50 & 0.51 & 0.51 & 0.50 & 0.44 &
0.36 & 0.34 & 0.33 & 0.29 & 0.25 & 0.25 & 0.26 & 0.25 & 0.22 & 0.25 & 0.37 & 0.48 & 0.62 \\

0.56 & 0.56 & 0.57 & 0.63 & 0.68 & 0.69 & 0.67 & 0.65 & 0.65 & 0.59 &
0.49 & 0.44 & 0.42 & 0.35 & 0.30 & 0.28 & 0.30 & 0.30 & 0.29 & 0.33 & 0.45 & 0.57 & 0.75 \\

0.69 & 0.66 & 0.69 & 0.79 & 0.85 & 0.83 & 0.78 & 0.75 & 0.76 & 0.73 &
0.64 & 0.54 & 0.47 & 0.41 & 0.34 & 0.30 & 0.31 & 0.35 & 0.37 & 0.40 & 0.49 & 0.65 & 0.85 \\

0.67 & 0.72 & 0.80 & 0.92 & 0.96 & 0.90 & 0.83 & 0.80 & 0.82 & 0.80 &
0.72 & 0.59 & 0.47 & 0.42 & 0.37 & 0.33 & 0.31 & 0.33 & 0.37 & 0.41 & 0.49 & 0.70 & 0.91 \\

0.64 & 0.76 & 0.85 & 0.95 & 0.98 & 0.94 & 0.89 & 0.85 & 0.84 & 0.81 &
0.75 & 0.63 & 0.50 & 0.40 & 0.36 & 0.35 & 0.32 & 0.31 & 0.35 & 0.41 & 0.50 & 0.69 & 0.88 \\

0.69 & 0.78 & 0.83 & 0.88 & 0.92 & 0.95 & 0.94 & 0.90 & 0.84 & 0.80 &
0.76 & 0.69 & 0.56 & 0.41 & 0.33 & 0.33 & 0.33 & 0.33 & 0.35 & 0.41 & 0.51 & 0.63 & 0.78 \\

0.81 & 0.81 & 0.82 & 0.86 & 0.92 & 0.97 & 0.95 & 0.90 & 0.87 & 0.83 &
0.78 & 0.71 & 0.61 & 0.47 & 0.36 & 0.31 & 0.30 & 0.31 & 0.34 & 0.40 & 0.50 & 0.57 & 0.61 \\

0.88 & 0.83 & 0.82 & 0.91 & 0.99 & 1.01 & 0.95 & 0.89 & 0.89 & 0.87 &
0.79 & 0.71 & 0.62 & 0.52 & 0.39 & 0.28 & 0.24 & 0.27 & 0.33 & 0.42 & 0.50 & 0.54 & 0.50 \\

0.80 & 0.81 & 0.86 & 0.99 & 1.07 & 1.07 & 1.00 & 0.94 & 0.92 & 0.87 &
0.78 & 0.68 & 0.58 & 0.47 & 0.34 & 0.23 & 0.20 & 0.26 & 0.37 & 0.48 & 0.56 & 0.59 & 0.56 \\

0.70 & 0.79 & 0.92 & 1.06 & 1.11 & 1.08 & 1.03 & 0.98 & 0.94 & 0.87 &
0.75 & 0.63 & 0.50 & 0.38 & 0.27 & 0.20 & 0.20 & 0.27 & 0.39 & 0.48 & 0.52 & 0.56 & 0.61 \\

0.64 & 0.77 & 0.95 & 1.07 & 1.08 & 1.02 & 0.99 & 0.98 & 0.94 & 0.84 &
0.70 & 0.56 & 0.43 & 0.32 & 0.26 & 0.25 & 0.24 & 0.27 & 0.36 & 0.39 & 0.38 & 0.44 & 0.56 \\

0.56 & 0.71 & 0.85 & 0.96 & 0.99 & 0.96 & 0.94 & 0.92 & 0.86 & 0.75 &
0.61 & 0.50 & 0.41 & 0.34 & 0.33 & 0.35 & 0.33 & 0.30 & 0.31 & 0.30 & 0.27 & 0.34 & 0.44 \\

0.36 & 0.54 & 0.75 & 0.82 & 0.87 & 0.90 & 0.88 & 0.83 & 0.76 & 0.67 &
0.55 & 0.43 & 0.37 & 0.41 & 0.46 & 0.44 & 0.33 & 0.24 & 0.25 & 0.24 & 0.20 & 0.29 & 0.35\\
\bottomrule
\end{tabular}
\label{tab:V_values1}
\end{table}

\begin{table}[H]
\centering
\caption{Post change AR-operator $\hat{\Psi}_2$ for the year 2022}
\scriptsize
\setlength{\tabcolsep}{3pt}
\renewcommand{\arraystretch}{1.1}

\begin{tabular}{*{23}{r}}
\toprule

-12.98 & -3.79 & -0.76 & 9.12 & 6.76 & -1.47 & 4.97 & 11.51 & 5.89 & 2.77 &
4.11 & -0.01 & -3.04 & 0.97 & 0.03 & -5.62 & -1.68 & 5.21 & 3.81 & 1.94 &
0.65 & -9.55 & 0.86 \\

7.99 & 1.28 & -1.09 & -3.64 & -1.56 & 2.05 & 0.06 & -2.35 & -0.34 & 2.38 &
3.57 & 3.31 & 2.37 & 1.70 & 2.02 & 1.77 & -2.24 & -5.86 & -4.13 & -1.01 &
0.46 & 1.09 & 4.48 \\

6.98 & 0.86 & -2.45 & -6.43 & -2.94 & 2.99 & -0.26 & -4.46 & -1.54 & 3.31 &
5.32 & 2.62 & -0.07 & 1.01 & 0.95 & -1.18 & -1.14 & -0.75 & -2.72 & -2.91 &
0.01 & 1.28 & -7.27 \\

3.94 & -1.24 & 2.92 & 0.59 & -0.93 & 0.31 & 0.94 & -1.04 & -4.21 & -1.64 &
4.07 & 3.20 & -0.41 & -0.71 & 0.13 & 0.68 & 1.60 & 1.99 & 0.82 & -0.92 &
-1.75 & -0.59 & -2.51 \\

-2.06 & -0.13 & 2.53 & 2.60 & 0.88 & -0.30 & 1.04 & 0.79 & -2.74 & -2.74 &
0.99 & 2.51 & 0.81 & -1.97 & -0.99 & 3.03 & 3.61 & 1.25 & -0.69 & -1.23 &
-0.89 & -1.09 & 1.50 \\

-5.54 & 2.34 & -2.11 & -0.77 & 1.05 & 0.85 & -0.10 & 0.11 & 1.58 & 0.59 &
-1.30 & 1.07 & 2.36 & -1.57 & -1.63 & 3.19 & 3.76 & -0.12 & -3.98 & -3.35 &
0.51 & -0.18 & 0.16 \\

1.22 & 0.99 & -0.75 & -0.44 & 0.03 & -0.28 & -0.93 & 0.11 & 2.79 & 2.17 &
-0.35 & 1.30 & 3.10 & 0.77 & -0.39 & 1.30 & 2.21 & 1.28 & -0.84 & -2.51 &
-2.22 & 0.50 & 0.32 \\

5.76 & -0.95 & 1.86 & 1.70 & -0.40 & -1.83 & -0.54 & 1.28 & 1.68 & 1.86 &
1.93 & 1.07 & 0.77 & 1.82 & 1.01 & -1.09 & 0.14 & 2.80 & 2.83 & -0.18 &
-2.64 & 1.80 & 1.88 \\

-0.29 & -0.45 & 0.41 & 1.74 & 0.36 & -1.49 & 0.79 & 2.68 & 1.06 & 1.26 &
2.43 & -1.02 & -3.41 & 0.08 & 0.71 & -2.57 & -1.34 & 1.85 & 1.59 & 1.23 &
2.64 & 3.53 & 3.36 \\

-4.27 & -0.24 & -1.74 & -0.87 & -0.14 & -0.01 & 0.98 & 2.41 & 3.06 & 1.45 &
-1.26 & -1.97 & -1.05 & -0.29 & -1.24 & -2.89 & -1.95 & -0.16 & -0.01 & 1.49 &
3.91 & 2.09 & 5.03 \\

-1.85 & -0.55 & -2.03 & -3.25 & -1.42 & 1.07 & -0.08 & 0.38 & 4.33 & 1.36 &
-4.98 & -0.80 & 4.73 & 1.04 & -2.71 & -2.30 & -1.89 & -1.56 & -0.18 & 0.29 &
-0.81 & -1.20 & 5.52 \\

1.67 & 0.62 & -0.15 & -1.09 & -0.63 & 0.21 & -0.77 & -1.73 & -1.23 & -1.18 &
-1.20 & 0.89 & 2.48 & 1.07 & -0.75 & -1.52 & -1.78 & -1.35 & -0.55 & -2.49 &
-5.31 & -1.23 & 2.46 \\

1.93 & 0.87 & 0.61 & 1.76 & 0.85 & -1.21 & -1.25 & -2.40 & -5.58 & -3.58 &
1.48 & 0.43 & -1.47 & 1.53 & 1.74 & -2.08 & -2.79 & -1.21 & -1.31 & -3.66 &
-5.03 & 0.28 & -1.98 \\

-1.35 & -1.55 & -1.66 & 0.93 & 0.74 & -1.69 & -1.92 & -1.13 & -1.43 & -3.03 &
-4.36 & -2.92 & 0.63 & 3.58 & 1.25 & -4.32 & -4.68 & -2.05 & -1.65 & -1.19 &
0.37 & 0.30 & -4.66 \\

0.14 & -2.39 & -1.92 & -0.88 & -1.12 & -2.14 & -1.94 & 0.29 & 2.62 & -1.19 &
-7.10 & -3.98 & 2.09 & 2.32 & -1.50 & -4.90 & -3.35 & -0.20 & -0.03 & 0.92 &
2.79 & -0.39 & -4.61 \\

4.39 & 0.40 & 0.75 & -1.58 & -2.95 & -2.30 & -1.03 & 0.64 & 1.95 & 0.39 &
-2.23 & -1.24 & 0.12 & -1.76 & -3.39 & -2.39 & 0.84 & 3.13 & 1.92 & 0.75 &
0.99 & -0.27 & -2.41 \\

-0.85 & 4.98 & 1.97 & -1.12 & -1.77 & -0.59 & 0.21 & -0.14 & -0.81 & 1.12 &
3.75 & 1.82 & -1.61 & -2.18 & -0.54 & 1.63 & 2.58 & 1.74 & -0.20 & -0.24 &
1.74 & 1.82 & 0.21 \\

-6.13 & 6.19 & 1.03 & -1.26 & -0.43 & 1.12 & 0.92 & -0.91 & -2.23 & 1.14 &
5.67 & 2.84 & -1.77 & -0.98 & 2.08 & 3.73 & 1.97 & -1.43 & -3.13 & -0.75 &
3.42 & 2.37 & 2.41 \\

0.53 & 1.65 & -0.07 & -3.15 & -2.44 & 0.64 & 0.81 & -0.45 & -0.61 & 0.63 &
1.95 & 1.27 & -0.70 & -2.05 & -0.90 & 1.55 & 1.27 & -0.89 & -1.90 & 0.03 &
2.31 & -1.32 & 3.62 \\

5.02 & 2.32 & 2.99 & -3.30 & -4.43 & 0.02 & 1.83 & 0.97 & -0.39 & -1.11 &
-0.78 & 0.22 & 0.23 & -1.52 & -2.42 & -1.29 & 0.81 & 2.17 & 1.82 & 1.91 &
1.76 & -2.36 & 2.54 \\

1.93 & 6.86 & 6.21 & -0.38 & -2.77 & 0.06 & 2.26 & 0.88 & -3.19 & -3.60 &
-0.56 & 0.21 & 0.13 & 1.37 & 0.79 & -0.97 & 0.91 & 4.05 & 4.28 & 3.18 &
2.16 & 0.97 & 0.21 \\

3.28 & -2.37 & -2.67 & 2.60 & 3.36 & -0.72 & -3.30 & -4.39 & -5.07 & -4.46 &
-2.89 & -1.74 & -0.54 & 1.36 & 3.11 & 3.54 & 1.78 & 0.40 & 1.55 & 1.04 &
-1.13 & 1.15 & -0.72 \\

6.97 & -1.39 & 2.18 & 0.56 & 1.27 & 3.16 & -0.51 & -3.48 & -1.16 & -0.52 &
-1.54 & 1.53 & 2.66 & -2.92 & -5.12 & -2.08 & -2.34 & -3.56 & -1.07 & -0.34 &
-3.41 & -6.05 & -15.51 \\

\bottomrule
\end{tabular}

\label{tab:V_values2}
\end{table}

%\textcolor{violet}{Lu, Zhaoying, and Yuanju Fang. "What Explains Bitcoin Volatility? Evidence from an Extended HAR Framework." International Journal of Financial Studies 14, no. 4 (2026): 81. This paper identifies Sept 11, 2022 as volatility break point. "This period coincides with heightened uncertainty in global financial markets, associated with rising monetary tightening expectations following unexpected inflation surprises in the United States. At the same time, cryptocurrency markets experienced heightened volatility surrounding Ethereum’s Merge upgrade in mid-September 2022."}
\subsection{LAND SURFACE MONTHLY AVERAGE TEMPERATURE}
We used the land-surface monthly average temperature series spanning 1753 to 2016, obtained from the Berkeley Earth dataset \url{http://berkeleyearth.org/data/}. These temperature observations are recorded in degrees Celsius and are expressed as anomalies relative to the baseline period from January 1951 to December 1980. To construct a functional time series, each year is represented by its 12 monthly average temperatures. This can be looked upon as 264 random curves on a grid of 12 points. Table~\eqref{tab:performance_comparison} presents the estimated change points under three distinct methods: the Bayesian approach of \citet{li2021bayesian}, the E-Divisive algorithm of \citet{matteson2014nonparametric}, and our proposed method. Notably, the former two approaches do not account for temporal dependence. Kernel density plot of $\hat{\sigma}_1$ and $\hat{\sigma}_2$
has been provided in Fig ~\eqref{var_gh}. Our proposed method explicitly identifies 1997 as a structural change point in the mean level of global temperatures. Evidence from geophysical research reveals that the 1997--1998 period was characterized by a string of record-breaking global temperatures, indicative of an accelerating warming rate due to radiative forcing from anthropogenic climate change, primarily driven by greenhouse gases (such as $\text{CO}_2$) and sulfate aerosols \citep{karl2000record}. We have obtained the change point in variance of the global temperatures in the year 1850. As documented by \cite{fagan2019little}, the period spanning 1300--1850 (commonly known as the Little Ice Age) was characterized by severe climate instability and high temporal volatility in global temperatures. The year 1850 marks the termination of this erratic climatic regime, delineating the highly variable pre-industrial climate swings from the subsequent modern temperature trajectory making it a natural structural change point in the variance of temperature series. \citet{li2021bayesian} also analyzed this dataset and compared their findings with the E-Divisive approach of \citet{matteson2014nonparametric}. However, both methods rely on the assumption of independent observations. The temporal dependence across the temperature curves is presented via the matrix $\hat{\Psi}$ presented in Table ~\eqref{Psi}. It is evident that the underlying data exhibits strong latent temporal dependence. Consequently, by accounting for this dependent structure, our proposed framework yields more reliable inference. We do not consider structural changes in the operator due to the high sparsity of the grid. Indeed, simulation results indicate that the method exhibits poor performance under such sparse conditions.\\

%\textcolor{violet}{Fagan, Brian. The Little Ice Age: how climate made history 1300-1850. Hachette UK, 2019.}

%\textcolor{violet}{Karl, Thomas R., Richard W. Knight, and Bruce Baker. "The record breaking global temperatures of 1997 and 1998: Evidence for an increase in the rate of global warming?." Geophysical Research Letters 27, no. 5 (2000): 719-722.}

%\begin{table}[H]
%\centering
%\caption{Performance comparison of the algorithm for the temperature data}
%\label{tab:performance_comparison}
%\renewcommand{\arraystretch}{1.9}
%\begin{tabular*}{\textwidth}{@{\extracolsep{\fill}}|c||c|c|c|c|}
%\hline
%\textbf{Year} & 
%\textbf{\makecell{CP\footnote{\citet{li2021bayesian}}}} & 
%\textbf{\makecell{CP\footnote{\citet{matteson2014nonparametric}}}}& 
%\textbf{\makecell{Mean CP}} & \textbf{\makecell{Variance CP}}\\
%\hline
%1753--2016 & \makecell{1761, 1767, 1779, 1801, 1812, 1821, 1839,\\ 1858, %1878, 1898, 1914, 1942, 1969, 1994} & 
%\makecell{1921, 1987,1851,1789,\\ 1819,1957,1888} & 
%1997 & 1850\\
%\hline
%\end{tabular*}
%\end{table}

%\footnotetext[1]{\citet{li2021bayesian}}
%\footnotetext[2]{\citet{matteson2014nonparametric}}

\begin{table}[H]
\centering
\caption{$\hat{\Psi}$ matrix representing latent dependence among the temperature curves} 
  \label{Psi} 
\small
\setlength{\tabcolsep}{5pt}
\begin{tabular}{rrrrrrrrrrrr}
\toprule

 2.36 &  2.95 &  3.29 &  3.43 &  3.12 &  2.65 &  2.29 &  1.97 &  1.25 & -0.10 & -1.36 & -2.51 \\
 2.55 &  2.67 &  2.85 &  2.90 &  2.65 &  2.31 &  2.03 &  1.58 &  0.84 & -0.12 & -1.10 & -2.25 \\
 2.48 &  2.07 &  2.14 &  2.30 &  1.95 &  1.67 &  1.78 &  1.41 &  0.64 & -0.06 & -0.93 & -1.85 \\
 1.34 &  1.04 &  1.40 &  1.59 &  1.42 &  1.47 &  1.86 &  1.51 &  0.62 & -0.10 & -0.65 & -0.77 \\
 0.23 &  0.21 &  0.75 &  1.16 &  1.09 &  1.18 &  1.68 &  1.55 &  0.86 &  0.20 & -0.17 & -0.52 \\
-0.53 & -0.33 &  0.24 &  0.94 &  1.07 &  1.18 &  1.60 &  1.58 &  1.14 &  0.71 &  0.31 & -0.36 \\
-1.02 & -0.66 & -0.04 &  0.81 &  1.31 &  1.63 &  1.90 &  1.70 &  1.34 &  1.18 &  0.68 &  0.41 \\
-1.44 & -0.64 &  0.11 &  0.97 &  1.53 &  1.84 &  2.00 &  1.93 &  1.76 &  1.60 &  1.11 &  0.82 \\
-1.54 & -0.32 &  0.55 &  1.29 &  1.71 &  1.92 &  2.08 &  2.16 &  2.10 &  1.83 &  1.34 &  0.58 \\
-1.16 &  0.06 &  0.96 &  1.58 &  2.01 &  2.31 &  2.45 &  2.30 &  2.04 &  1.76 &  1.07 & -0.05 \\
-0.89 &  0.08 &  1.15 &  1.88 &  2.55 &  2.85 &  2.65 &  2.41 &  2.11 &  1.50 &  0.63 & -0.63 \\
-0.40 &  0.35 &  1.11 &  1.86 &  2.87 &  3.44 &  3.15 &  2.63 &  2.09 &  1.42 &  0.49 & -1.01 \\
\bottomrule
\end{tabular}
\end{table}

\begin{figure}[H]
\centering
\includegraphics[width=0.90\textwidth]{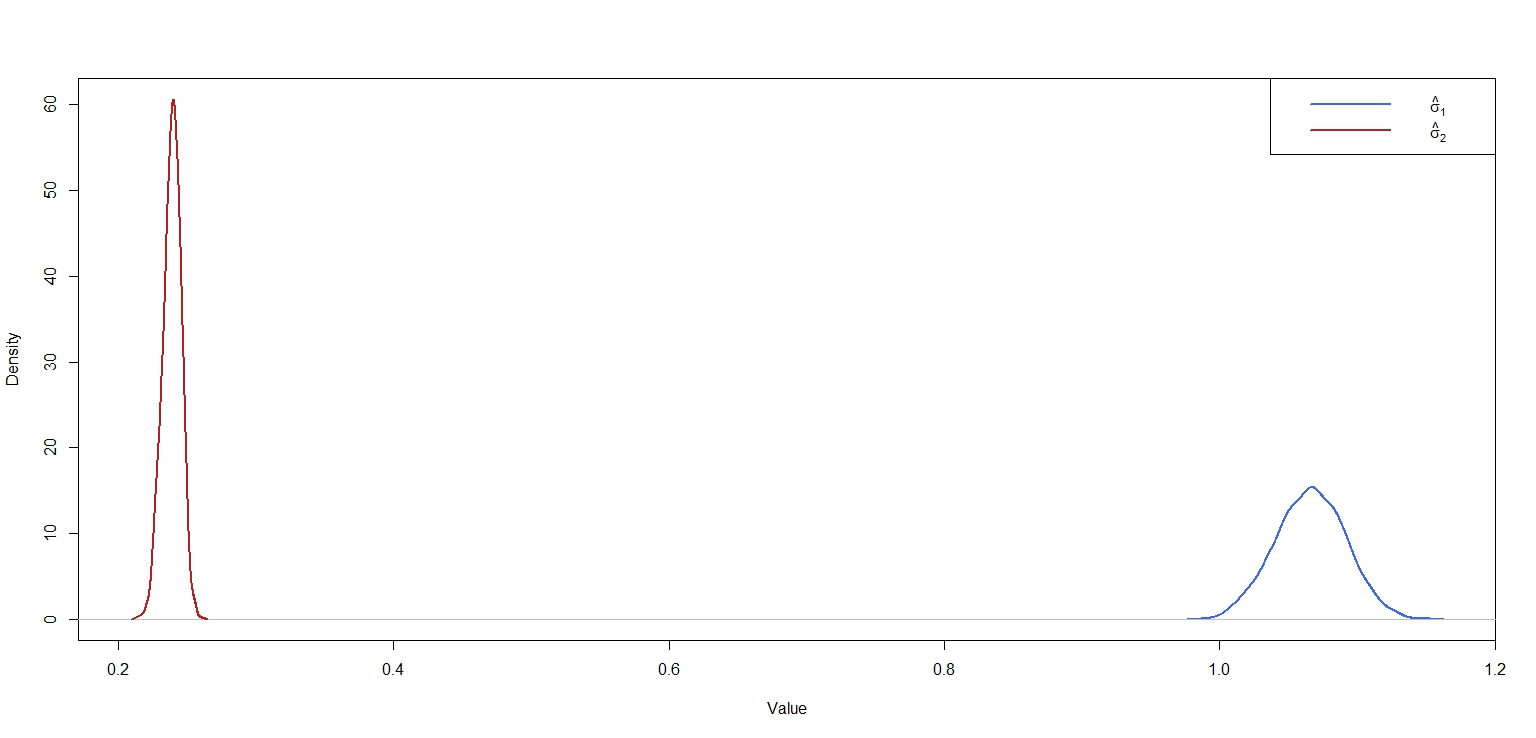}
\caption{Kernel density estimates of pre-change volatility ($\hat{\sigma}_1$) and post-change volatility ($\hat{\sigma}_2$) for the year 1850 across iterations.}
{\label{var_gh}}
\end{figure}

\begin{figure}[H]
\centering
\includegraphics[width=0.90\textwidth]{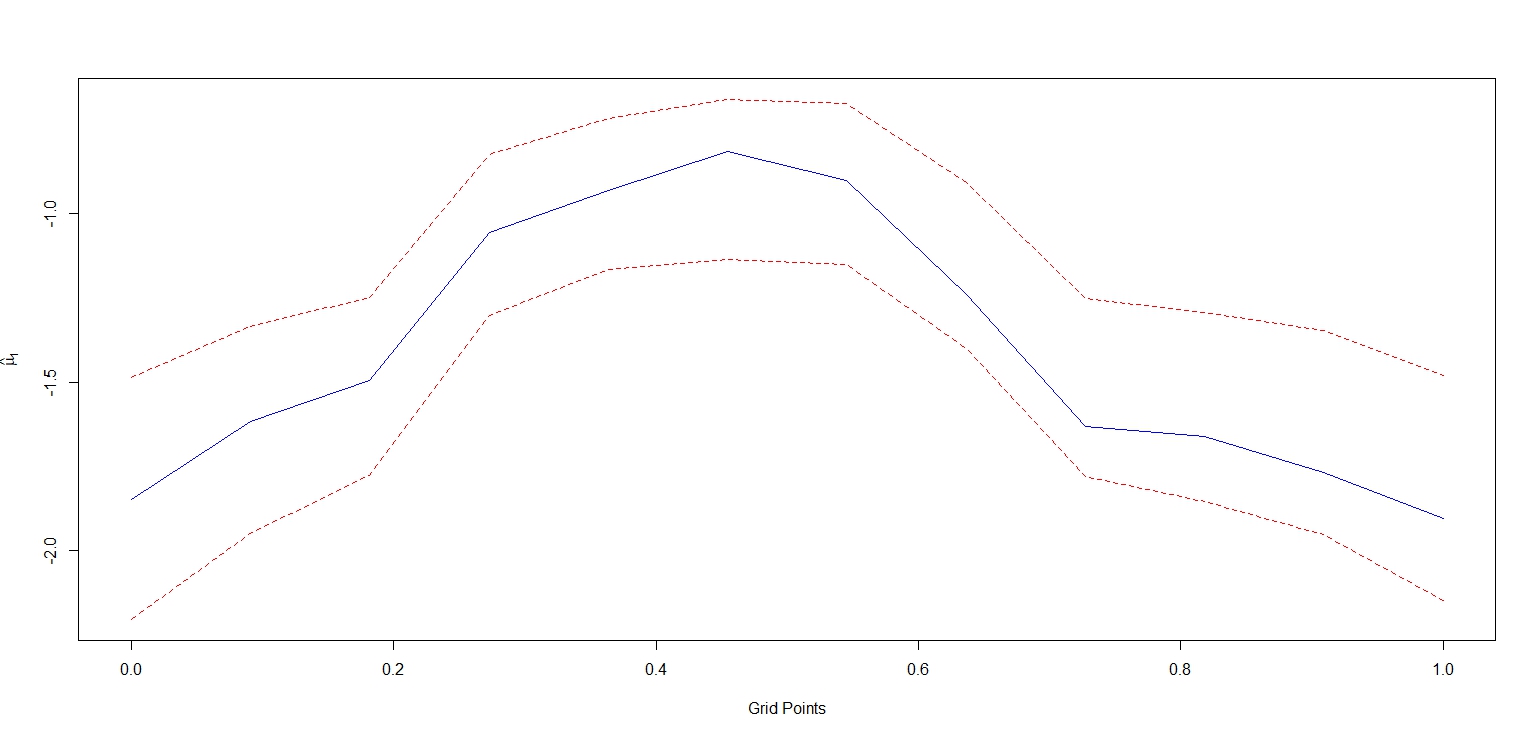}
\caption{95\% HPD credible interval containing $\hat{\mu}_1$ for temperature data}
{\label{var_gh}}
\end{figure}

\begin{figure}[H]
\centering
\includegraphics[width=0.90\textwidth]{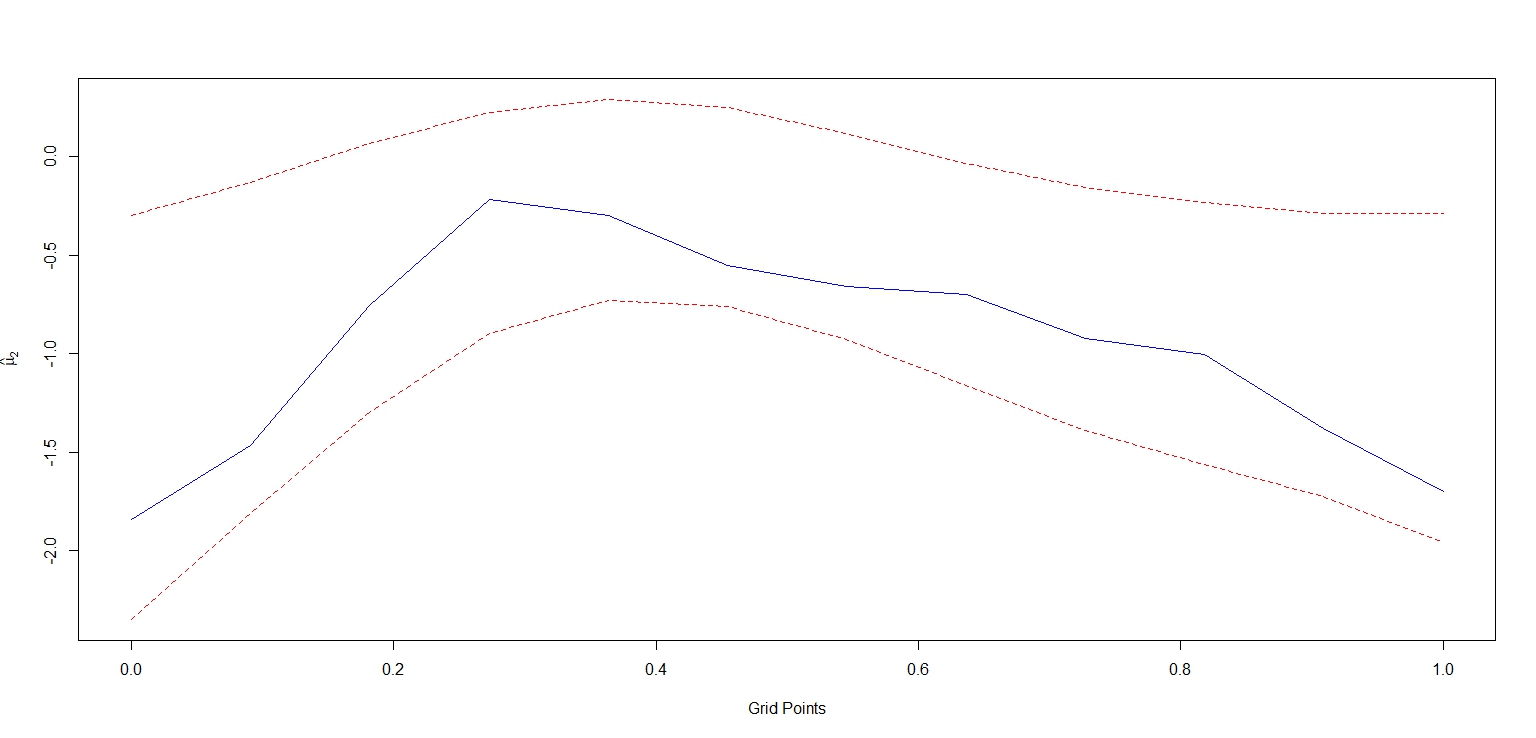}
\caption{95\% HPD credible interval containing $\hat{\mu}_2$ for temperature data}
{\label{var_gh}}
\end{figure}

%\begin{figure}[H]
% \centering
 % \begin{subfigure}[b]{0.80\textwidth}
 % \centering
 % \includegraphics[width=\textwidth]{ghoshaldata_beforeCP.jpeg}
 % \caption{Before changepoint}
% \end{subfigure}
 % \hfill
 % \begin{subfigure}[b]{0.80\textwidth}
 % \centering
 % \includegraphics[width=\textwidth]{ghoshaldata_afterCP.jpeg}
 % \caption{After changepoint}
 % \end{subfigure}
 % \caption{Comparison of data behavior around the structural shift.}
 % \label{fig2}
%\end{figure}

\subsection{NIFTY 50}
The onset of the COVID-19 pandemic in early 2020 represented one of the most abrupt and globally synchronised economic shocks in modern history, fundamentally altering the landscape for the Indian equity market. The NIFTY 50, which serves as the premier benchmark index reflecting the performance of the largest and most liquid Indian companies, provides a sharp lens  to examine this situation. It has been already pointed out in \cite{barik2020impact} that there has been a shift in aviation stock volatility around the year 2020.

\begin{figure}[H]
\centering
\includegraphics[width=0.50\textwidth]{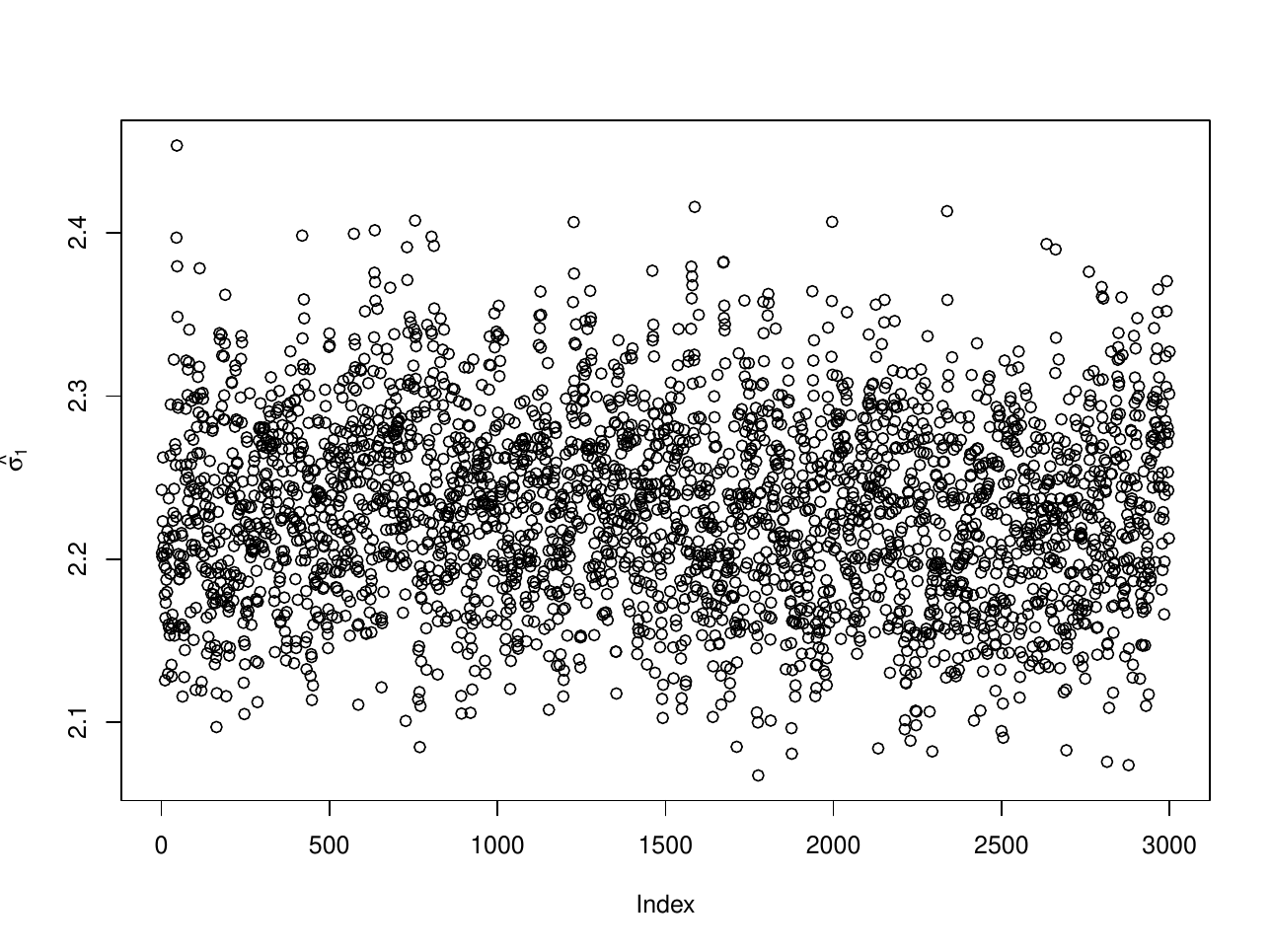}
\includegraphics[width=0.50\textwidth]{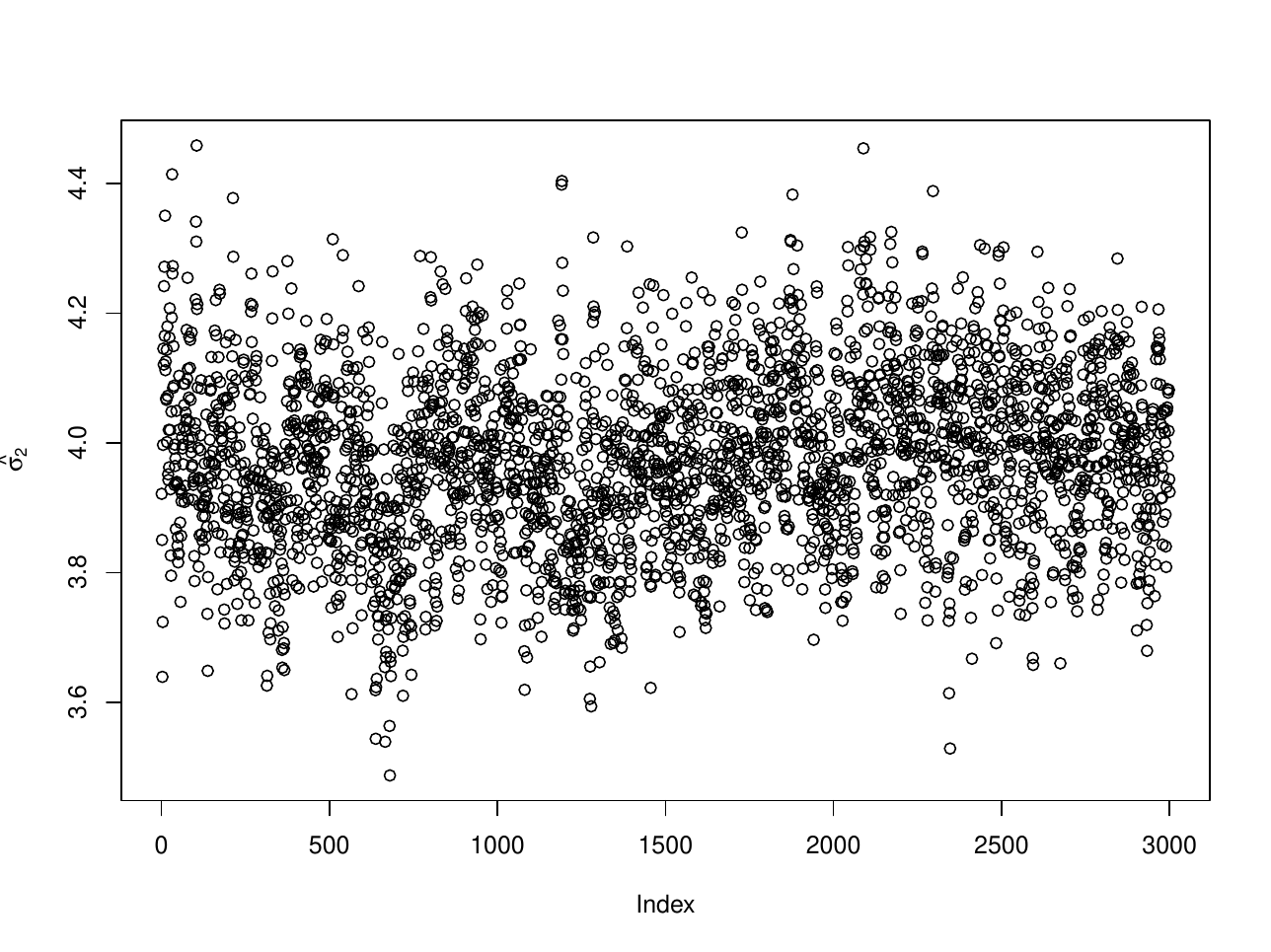}
\caption{Plots of pre-change volatility ($\hat{\sigma}_1$) and post-change volatility ($\hat{\sigma}_2$) across iterations in the scale of $10^3$}{\label{fig1}}
\end{figure}

The objective of this investigation is to examine whether our proposed methodology can capture the volatility structural break in the NIFTY 50 index induced by the COVID-19 pandemic. Given the empirical consensus regarding the severe volatility shock brought by the pandemic, this study evaluates the model's sensitivity and precision in detecting these exogenous market dislocations.
Consider the high-frequency intra-day data for the NIFTY 50 index, specifically the hourly log returns of the closing prices measured between 9:00 AM and 3:00 PM across a two-year period from January 1, 2019 to December 31, 2020. The change in volatility occurred on February 28, 2020, which aligns precisely with the primary change point identified by the E-Divisive method of \citep{matteson2014nonparametric}. Plots of the estimated values of $\hat{\sigma}_1$ and $\hat{\sigma}_2$ over iterations are provided in Figure \ref{fig1}.

From this analysis, we observe that after the onset of COVID-19, NIFTY 50 stocks became more volatile.
This increased volatility can be accounted for by unprecedented global uncertainty, massive capital outflows, and a structural shift in how the market participants processed the information. 
\cite{zhang2020financial} highlights that global financial market risks increased substantially in response to the COVID--19 pandemic. Individual stock market reactions were closely associated with the severity of the outbreak across countries. The unprecedented uncertainty surrounding the pandemic, together with its significant economic consequences, resulted in heightened market volatility and increased unpredictability. \cite{okorie2021stock} documented that the COVID--19 pandemic intensified stock market volatility and generated significant short-run volatility contagion across global financial markets. Hence, our results are in line with the existing literature.\\

\noindent
%\textcolor{violet}{Add more references to the NIFTY example. One such is Zhang, Dayong, Min Hu, and Qiang Ji. "Financial markets under the global pandemic of COVID-19." Finance research letters 36 (2020): 101528. Quoting from hte paper: The present results show that global financial market risks have increased substantially in response to the pandemic. Individual stock market reactions are clearly linked to the severity of the outbreak in each country. The great uncertainty of the pandemic and its associated economic losses has caused markets to become highly volatile and unpredictable.}

\section{CONCLUSION}{\label{conc}}
In this article, we have investigated several techniques based on both single and multi-parameter distortion setups in a sequence of random functions. The proposed algorithms rely on an offline scheme, basically bisecting the support without any sequential update. %Simultaneous change point detection (SCPD) is vital because it treats the system as a holistic entity.
By aggregating weak signals across an unknown, dynamically selected subset of coordinates, simultaneous change point detection (SCPD) methodology achieves higher sensitivity, allowing it to detect subtle, coordinated systemic shifts that independent univariate methods completely miss. Furthermore, it offers robust performance by maintaining a strictly controlled global false alarm rate regardless of the number of monitored dimensions, preventing the statistical explosion of false positives. Finally, SCPD provides exceptional agnosticism, requiring zero prior knowledge of the specific failure mode or which parameters will shift, making it uniquely viable for unpredictable, complex real-world events. So conditional on no change in other parameters, SCPD gives a similar result as the individual method for the specific parameter. This instance has been clearly depicted in Table ~\eqref{tab:research_data}. In general, change point detection techniques designed for independent random curves perform poorly or fail entirely when structural dependencies are present in the data. Conversely, we demonstrate that our proposed framework not only robustly handles these complex dependencies, but also maintains equivalent efficacy and performance when applied to strictly independent data structures. This notion has been clearly depicted in Table ~\eqref{var_gh} where as a support of our procedure, an AR-operator matrix depicting the latent temporal dependence has been provided in Table ~\eqref{Psi}. The performance of the proposed method has been rigorously evaluated and validated on the NIFTY 50 dataset. The obtained results demonstrate that the proposed approach is capable of effectively capturing the underlying patterns and dynamics present in the financial time-series data. In particular, the observed behavior of the model is consistent with the characteristics of the physical phenomenon under consideration, thereby providing a meaningful interpretation of the results from both theoretical and practical perspectives. Furthermore, the findings are in agreement with the trends and observations reported in the existing literature, which further supports the reliability and robustness of the proposed methodology. The consistency between the empirical results, the underlying physical principles, and previously established findings provides strong evidence that the proposed method is well suited for analyzing and modeling the dynamics represented in the NIFTY 50 dataset.
%Additionally, a retrospective examination of key global events reveals critical dynamic shifts that our methodology effectively detects. To address potential distortions driven by external factors, new approaches can be developed beyond rigid model constraints, isolating such distortions directly via the structural association of the link functions. Future directions of work may include investigations in the covariate-augmented DLM and generalised DLM(GDLM). Rather than considering at most one changepoint, new methodologies can be developed for the detection of multiple changepoints. 
\bibliography{References}
\section{Appendix}
\begin{definition}
 A Functional Autoregressive Moving Average Process of order $(p, q)$, (FARMA$(p, q))$, is defined by
\begin{equation*}
\Phi(B)(\alpha_t ) = \Theta(B)\epsilon_t;
\end{equation*}
where, $\Phi(B)$ is a polynomial of order $p$ in the backshift operator $B$, such that 
\begin{equation*}
\Phi(B)\alpha_t = (1 - \phi_1 B - \phi_2 B^2 - \dots - \phi_p B^p)\alpha_t = \alpha_t - \sum_{\ell=1}^{p} \phi_\ell (\alpha_{t-\ell})
\end{equation*}
where, $\{\phi_\ell\}_{\ell=1}^p$ are bounded linear operators on $L^{2}$. Similarly, $\Theta(B)$ is also a polynomial of order $q$ in the backshift operator $B$, where, $\{\theta_\ell\}_{\ell=1}^q$ are bounded linear operators on $L^{2}$.
\end{definition}

\subsection{ROBUST KALMAN FILTER }{\label{kalman}}
The Kalman filter is a recursive state estimation algorithm designed for linear dynamic systems observed through noisy measurements. Introduced by \cite{kalman1960new}, it provides optimal estimates of the latent state variables under the assumptions of linearity, Gaussian disturbances, and known system parameters. The algorithm operates sequentially through two stages: a prediction step, in which the current state estimate is propagated forward using the system dynamics, and an update step, where the prediction is corrected using newly observed data. Owing to its recursive structure, the Kalman filter is computationally efficient and well suited for real-time applications. A particularly simple and practical formulation of the filtering and smoothing recursions was developed by \cite{durbin2002simple}, making the methodology accessible for a wide range of time series and state-space modeling applications.

Despite its optimality under Gaussian assumptions, the classical Kalman filter is known to be sensitive to anomalous observations. In practice, measurement errors or system disturbances may occasionally produce observations that deviate substantially from the assumed model. Such observations, commonly referred to as outliers, can adversely affect the filtering process and lead to inaccurate state estimates. Of particular concern are propagating outliers, whose effects are not confined to a single time point but may persist and contaminate subsequent state estimates through the recursive nature of the algorithm. These outliers can be classified as exogenous, arising from contamination in the observation process, or endogenous, originating from disturbances in the state evolution equation itself.

To mitigate the influence of such aberrant observations, several robust modifications of the Kalman filter have been proposed. One prominent approach, adopted in this paper, is based on the concept of \emph{Huberization}, wherein the correction term in the update step is bounded using Huber's robust loss function. This modification reduces the influence of extreme innovations while preserving the efficiency of the classical Kalman filter for observations that are consistent with the underlying model. Consequently, the robustified Kalman filter provides stable and reliable estimates of the latent state vectors, $\{\alpha_t\}_{t=1}^{T}$, even in the presence of substantial contamination. In the proposed Bayesian framework, the robust Kalman filter is employed at every iteration of the MCMC algorithm to mitigate the effects of observation misclassification, thereby yielding more reliable estimates of the latent state vectors. A comprehensive treatment of robust state-space methods and Huberized Kalman filtering is provided by \cite{ruckdeschel2014robust}.

\end{document}